\documentclass[aps,showpacs,pre,preprint,amsfonts,floats,graphicx]{revtex4}

\usepackage{graphicx}
\begin{document}

\bibliographystyle{amsplain}

\title{Nonintegrable Schr\"odinger Discrete Breathers }
\author{J. G\'omez-Garde\~nes}

\email[Corresponding author: ]{gardenes@unizar.es}
\affiliation{Dpt. de Teor\'{\i}a y Simulaci\'on de Sistemas 
Complejos. Instituto de Ciencia de Materiales de Arag\'on. 
C.S.I.C.-Universidad de Zaragoza, 5009~Zaragoza, Spain.}

\affiliation{Dpto.\ de F\'{\i}sica de la Materia Condensada e
Instituto de Biocomputaci\'on y F\'{\i}sica de los Sistemas Complejos 
(BIFI), Universidad de Zaragoza, 50009~Zaragoza, Spain}

\author{L.M. Flor\'{\i}a} 

\affiliation{Dpt. de Teor\'{\i}a y Simulaci\'on de Sistemas 
Complejos. Instituto de Ciencia de Materiales de Arag\'on. 
C.S.I.C.-Universidad de Zaragoza, 5009~Zaragoza, Spain.}

\affiliation{Dpto.\ de F\'{\i}sica de la Materia Condensada e
Instituto de Biocomputaci\'on y F\'{\i}sica de los Sistemas Complejos 
(BIFI), Universidad de Zaragoza, 50009~Zaragoza, Spain}

\author{M. Peyrard}

\affiliation{ Laboratoire de Physique. Ecole Normale
Sup\'erieure de Lyon, 69364~Lyon, France}

\author{A.R. Bishop}
\affiliation{Theoretical Division and Center for Nonlinear
Studyes, Los Alamos National Laboratory, Los Alamos, New Mexico~87545} 

\date{\today}

\begin{abstract}
In an extensive numerical investigation of nonintegrable
translational motion of discrete breathers in nonlinear
Schr\"odinger lattices, we have used a regularized Newton algorithm
to continue these solutions from the limit of the integrable Ablowitz-Ladik
lattice. These solutions are shown to be a
superposition of a localized moving core and an excited extended
state (background) to which the localized moving pulse is
spatially asymptotic. The background is a linear combination of
small amplitude nonlinear resonant plane waves and it plays an
essential role in the energy balance governing the translational
motion of the localized core. Perturbative collective variable
theory predictions are critically analyzed in the light of the
numerical results.
\end{abstract}


\maketitle 

{\bf Discrete breathers are spatially localized, time periodic solutions of
homogeneous nonlinear lattices, which have been recently observed in
experiments on a variety of physical systems (magnetic solids, arrays
of Josephson junctions, coupled optical waveguides and photonic
crystals). Though many of the properties of discrete breathers are
today well characterized, the question on their mobility remains under
controversy, due to the radiative losses unavoidably associated to the
traslational motion of the localized pulse in generic (nonintegrable)
systems. We address here this problem in an important class of
nonlinear lattices: {\em the discretizations of the Nonlinear
Schr\"odinger Equation}. Our results show that exact mobile breather
solutions ride over an extended excited state of the lattice, which we
fully characterize. Moreover this background p
lays an essential role
in the energy balance required for exact nonintegrable mobility. 
}

\section{Introduction} 
\label{sec:1}

Nonlinear lattices have become the subject of a considerable
multidisciplinary interest, with applications in physics
subdisciplines as diverse as biophysics (myelinated nerve fibers
\cite{Scott}, DNA \cite{Peyrard_DNA}, biopolymer chains
\cite{Mingaleev}), nonlinear optical devices (photonic crystals
\cite{McGurn} and waveguides \cite{KivsharAgrawal,Christo}), and
Josephson effect \cite{Leggett} (superconducting devices \cite{Trias,Binder},
Bose-Einstein condensates \cite{Strecker,Cataliotti,Smerzi}), 
among others. From a theoretical perspective they have been 
progressively recognized not as mere discretizations 
(unavoidable for numerical computations) of nonlinear continuum 
field equations, but as a target of interest in their own right, 
due to the distinctive features associated with {\em discreteness}, 
whose relevance to experimental features have been largely established.

More specifically, among the variety of observable nonlinear
behaviors of lattice dynamics, the phenomenon of (discrete
breathers) nonlinear localization in lattices
\cite{Campbell_PhysToday} has received 
attention in both experimental and theoretical research during the
last several years. (Nontopological) discrete breathers are exact
spatially localized, time-periodic solutions. Due to discreteness
the plane wave spectra are bounded, thus making posible the
absence of multi-harmonic resonances of the exact discrete
breather solution with extended modes. The combination of
nonlinearity and discreteness is sufficient for the physical
existence of discrete breathers, DB's for short, resulting in 
its generality and broad interest. The reader may 
find in \cite{KivsharFlachChaos} a recent multidisciplinar 
survey of current research on the subject.

Our primary concern here is the issue of DB's mobility. The
translational motion of discrete breathers introduces a new time
scale (the inverse velocity), so generically a moving
breather excites resonances with plane wave band spectra. This
fact poses no problem to the persistence of localization when the
lattice dynamics is governed by power balance (forced and damped
lattices \cite{Martinez}): the emitted power is exactly
compensated by the input from the homogeneous external force
field, during stationary breather motion. However, for generic
(nonintegrable) hamiltonian lattices, the radiative losses would
tend to delocalize energy, and some energy compensating mechanism
is needed in order to sustain exact stationary states of breather
translational motion. From the (particle) perspective of
collective variables theory, the localized breather experiences a
periodic Peierls-Nabarro potential function of its position, so
that the motion of the localized field oscillation over this
landscape should be expected to induce the emission of radiation
at the expense of translational (and/or internal) breather kinetic
energy, which thus would unavoidably decay on time.

To address the problem, a reasonable strategy is to use precise 
numerical techniques on adequately general models, with the hope 
that they may pave the way to further physical (and mathematical) 
insights. Our chosen model belongs to an important class of 
nonlinear lattice models: the discretizations of the continuum 
nonlinear Schr\"odinger equation, here referred to as 
Nonlinear Schr\"odinger (NLS) lattices. First of all, they 
are ubiquitous in models for polaronic effects in 
condensed matter, nonlinear optical technologies, and
the physics of Bose condensate lattices and superconducting
devices, where nonlinear localization is currently studied. 
Second (a very convenient technical advantage), this class 
contains an integrable limit (the Ablowitz-Ladik lattice, see 
below) having exact moving discrete breathers, wherefrom perturbation
(collective variable) theories have been developed in support of
exact (or very approximate, in the least) nonintegrable mobile 
DB's. This allows a detailed theoretical analysis of the
numerical results, as well as eventual feedbacks for useful (and
currently used) theoretical concepts and perspectives. In the
subsection \ref{NLSbreathers} of this introduction we present 
the (three-parameter) NLS lattice that we have studied, 
originally introduced by Salerno \cite{Salerno}.

The numerical techniques employed are summarily introduced in
section II. We stress here the unbiased character of this
numerical procedure which, unlike other techniques, is not based
on ansatze on the expected functional form of the exact solution
sought. In essence, the procedure uses a regularized Newton
continuation method for operator fixed points, and it only
requires a good starting set of parameter values where the exact
solution is known. In our case this is provided by the integrable
Ablowitz-Ladik limit of the NLS lattice, from which adiabatic
continuation of the two-parameter (core frequency $\omega_b$ and
velocity $v_b$) family of moving Schr\"odinger breathers is
performed.

\begin{figure}[!tbh]
\begin{tabular}{cc}
\centerline{\resizebox{8.cm}{!}{%
\includegraphics[angle=-0]{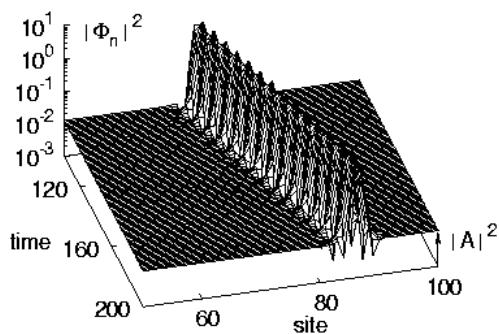}
}}
\end{tabular}
\caption{Time evolution of $|\Phi_{n}|^2$ profile of
a mobile discrete Schr\"odinger breather. The frequency of the 
solution is $\omega_{b}=5.050$ and the velocity is $v_{b}=0.804$. 
Note that the background is composed by a single plane wave with 
amplitude $A$. The nonintegrable parameter of eq. (\ref{Salerno}) 
is $\nu=0.2$. 
} \label{fig:Mobile}
\end{figure}

The main numerical facts are shown in section III. The numerical
solutions are found to be (up to numerical precision) the 
superposition of a traveling exponentially localized oscillation
(the {\em core}), and an extended background, which is
a linear superposition of finite amplitude nonlinear plane waves
$A\exp[i(kn - \omega t)]$ (see Fig. \ref{fig:Mobile}). The amplitudes of these
resonant nonlinear plane waves are observed to differ typically
by orders of magnitude, so that only a small number of them are
relevant for most practical purposes. They fit well simple
theoretical (thermodynamic limit) predictions based on discrete
symmetries requirements. Contrary to the exact immobile breather
solution (space-homoclinic and time-periodic orbit), which
asymptotically connects the rest state (vacuum or ground state) of
the lattice with itself, each exact mobile localized solution is
instead homoclinic to a specific lattice state of extended
radiation. In other words, exact stationary mobility of discrete
breathers requires an extended excited state of the lattice. A
preliminary account of some of the numerical results of this
section were reported in \cite{Gomez}.

In section IV we analyze the numerical results in the light of
collective variable theories, correlating them with the main
theoretical predictions of this successful (however incomplete)
physical perspective. In particular, the existence of
Peierls-Nabarro barriers to translational core motion is
confirmed, and its subtle relation to the background amplitude is
discussed. We present also numerical confirmation of the existence
of exact oscillating anchored breathers, whose extended background
is much smaller than those of traveling discrete breathers of the
same internal frequency $\omega_b$. Along with the discussion in this
section, a physical interpretation of the role of the interaction
background-core in the energy balance emerges, paving the way to a
satisfactory integration of the results into a collective
variable theory.

Finally, in section V, after summarizing the main conclusions
drawn on discrete Schr\"odinger breather mobility, we briefly trace
some interesting open questions for further research,
notably the approach to irrational breather time scales ratios, the
study of multibreather solutions (two-breather collision
processes, trains of moving breathers), and the coupling to both
thermal and nonthermal ({\em e.g.} elastic) degrees of freedom,
where the numerical tools and results presented here can find
further applications.

\subsection{NLS lattices}
\label{NLSbreathers}

The standard discrete nonlinear Schr\"odinger (DNLS) equation
\cite{Scott,Eilbeck} is the simplest discretization
of the one-dimensional continuous Schr\"odinger equation with cubic
nonlinearity in the interaction term, {\em i.e.},

\begin{equation}
{\mbox i} \dot{\Phi}_n= -C(\Phi_{n+1} + \Phi_{n-1}) - \gamma
|\Phi_n|^2 \Phi_n\;. \label{DNLS}\end{equation}

In this expression $\Phi_n(t)$ is a complex probability amplitude,
the parameter $C$ amounts the nearest neighbor coupling, and
$\gamma$ is the strength of the nonlinearity. The self-focussing
effect of local nonlinearity balanced by the opposite effect of
the dispersive coupling makes possible the existence of localized
periodic solutions (breathers) of the discrete field, where the profile of
$|\Phi_n|$ decays exponentially away from the localization
center:
\begin{equation}
\Phi_{n}(t)=|\Phi_n|\exp[{\mbox i}\omega_{b}(t))]\;.
\label{staticbreather}
\end{equation}
In the uncoupled limit $C \rightarrow 0$ of the DNLS equation,
also known as the anti-integrable or anti-continuous limit, discrete
breathers can be easily constructed by selecting a periodic
oscillation $\Phi_{n_0}(t)$ of frequency $\omega_b$ at site $n_0$
and $\Phi_n=0$ for $n \neq n_0$. These solutions can be uniquely
continued to nonzero values of the coupling $C$, and constitute
the one-parameter family of immobile on-site breathers of the
DNLS equation.

Unfortunately the continuation from the uncoupled limit does not
provide solutions where the localization center moves along the
lattice with velocity $v_{b}$, {\em i.e}, mobile discrete breathers. 
On the other hand, there is an integrable lattice as a limit of 
the nonlinear Schr\"odinger class that posseses this type of 
mobile solution. That is the one discovered by Ablowitz and 
Ladik in \cite{AL}:
\begin{equation}
{\mbox i} \dot{\Phi}_n= -C(\Phi_{n+1} + \Phi_{n-1})\left[ 1 +
\frac{\gamma}{2}|\Phi_n|^2 \right]\;,   
\label{Ablowitz}
\end{equation}
where, again, $C$ and $\gamma$ account for the strength of the coupling
and the nonlinearity, respectively. The integrable Ablowitz-Ladik
equation, A-L for short, possesses a two-parameter family of exact 
moving breather solutions of the form
\begin{eqnarray}
\Phi_n (t) &=&\sqrt{\frac{2}{\gamma}} \sinh \beta \; \mbox{sech}
[\beta (n-x_0(t))]\times \nonumber
\\
&\exp& [{\mbox i}(\alpha (n-x_0(t)) +\Omega(t))]\;.
\label{A-Lbreather}
\end{eqnarray}
The two parameters of this breather family can be chosen to be the
breather frequency $\omega_b$ and velocity $v_b$,
\begin{eqnarray}
v_b&=&\dot{x}_0=\frac{2\sinh \beta \sin \alpha}{\beta}
\\
\omega_b&=&\dot{\Omega}=2\cosh \beta \cos \alpha \;+\; \alpha
v_b\;, \label{omegav}
\end{eqnarray}
where $-\pi \leq \alpha \leq \pi$ and $0 < \beta < \infty$. The
A-L moving breather (instantaneous) profile interpolates between
the rest state $\Phi_n =0$ of the lattice (at $n \rightarrow \pm
\infty$) in an exponentially localized region around $x_0(t)$,
while traveling with velocity $v_b$.

The connection between the integrable (though physically limited)
A-L equation and the physically relevant (though nonintegrable) DNLS
equation is provided by the model originally introduced by Salerno 
in \cite{Salerno},
\begin{equation}
{\mbox i} \dot{\Phi}_n= -(\Phi_{n+1} + \Phi_{n-1})\left[ C + \mu
|\Phi_n|^2 \right] - 2 \nu \Phi_n |\Phi_n|^2\;.
\label{Salerno}
\end{equation} 
This lattice provides a Hamiltonian interpolation between the 
standard DNLS equation (\ref{DNLS}), for $\mu= 0$ and 
$\nu = \gamma/2$, and the integrable A-L lattice when 
$\mu = \gamma/2$ and $\nu =0$. The Hamiltonian of the
Salerno equation is given by
\begin{eqnarray}
{\cal H}= &-&C\sum_{n}(\Phi_{n}\overline{\Phi}_{n+1} +
\overline{\Phi}_{n}\Phi_{n+1}) -2\frac{\nu}{\mu}\sum_{n}
|\Phi_n|^2 \nonumber
\\
&+&2\frac{\nu}{\mu^2}\sum_{n}\ln(1+\mu|\Phi_n|^2)\;,
\label{Salerno_Ham}
\end{eqnarray}
which contains the A-L and DNLS Hamiltonian for the above limits. In
addition to the Hamiltonian, this equation possesses, for any value of the
parameters, the following conserved norm:
\begin{equation}
N = \frac{1}{\mu}\sum_{n}\ln(1+\mu|\Phi_n|^2)\;.
\label{norm}
\end{equation}
In the following we will set the value of $\gamma=2$ (as usual) and consider the
coupling strenght $C=1$ in eq. (\ref{Salerno}).

The continuation of the family of mobile discrete breathers from the
A-L integrable limit allows numerical observations of the interplay 
between the integrable term, weighted by the parameter $\mu$, and the 
nonintegrability, weigthed by $\nu$.

\section{Discrete Breathers numerics}
\label{sec:II}

We introduce here the numerical techniques that we have used. 
As a whole, one could refer to them as the (SVD-)
regularized Newton method. They do not constitute a novel method 
in "discrete Breather numerics", as they have been
already used, {\em e.g.} in \cite{CretegnyAubry} 
to refine moving breathers of Klein-Gordon lattices obtained by 
other numerical means (see, by contrast, \cite{Kladko}). 
From the methodological side, what is novel here is the
systematic use of them in the investigation of the family of
moving Schr\"odinger breathers reported below in \ref{sec:Mobile}.

To some extent, the presentation here is self-contained.
First, in \ref{sub:resonant} we introduce the notion of ($p$, $q$)
resonant solution, providing some illustrative examples. The (SVD)
regularized Newton algorithm is presented in \ref{Newton}, and
finally in \ref{Floquet.Ana} we briefly explain the basics of
Floquet stability analysis.

\subsection{Discrete space-time symmetries: ($p$, $q$) resonant states}
\label{sub:resonant}

If a frequency $\omega_b = 2\pi/T_b$ is given, we will say that a
solution $\Phi =\{\Phi_n(t)\}$ is ($p$, $q$) {\em resonant} with
respect to the reference frequency $\omega_b$, if the following
condition holds, for all $n$ and $t$:
\begin{equation}
\Phi_n(t) = \Phi_{n+p}(t+qT_b)\;. \label{resonance}
\end{equation}

After $q$ $T_b$-periods, these solutions repeat the same profile
but displaced by $p$ lattice sites. In more technical terms, these
($p$, $q$) resonant solutions are fixed points $\Phi$ of the
operator
\begin{eqnarray}
{\cal L}^{p} {\cal T}^{q}&=& {\cal M}
\\
({\cal M} - {\cal I}) \; \Phi &=& 0\;, \label{operators}
\end{eqnarray}
where ${\cal L}$ and ${\cal T}$ are, respectively, the lattice
translation and the $T_b$-time evolution operator
\begin{eqnarray}
{\cal L} \{\Phi_n(t)\} &=& \{\Phi_{n+1}(t)\}
\\
{\cal T} \{\Phi_n(t)\} &=& \{\Phi_n(t+T_b)\}\;.
\end{eqnarray}

\begin{figure}[!tbh]
\begin{tabular}{cc}
\centerline{\resizebox{8cm}{!}{%
\includegraphics[angle=-90]{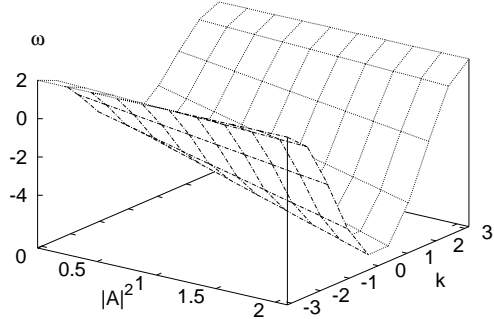}
}}
\end{tabular}
\caption{Plot of the nonlinear dispersion relation surface of
nonlinear plane waves, eq. (\ref{dispersion}), as a function of 
the amplitude $A$ and the wave number $k$ of the plane wave. 
The values of $\mu$ and $\nu$ are fixed to $0.5$.} 
\label{fig:surface}
\end{figure}

We now consider some examples of ($p$, $q$) resonant solutions
with respect to the frequency $\omega_b$; the first example is
simply provided by the family of plane wave solutions of eq.
(\ref{Salerno}):
\begin{equation}
 \Phi_{n}(t)=A \exp [{\mbox i}(k n -\omega t)]\;.
\label{plane-wave}
\end{equation}
It is easily seen, by inserting (\ref{plane-wave}) in eq.
(\ref{Salerno}), that the values of $\omega$, $k$ and $|A|$ define
a surface in the three-dimensional space, the nonlinear {\em
dispersion relation surface} $\omega(k,A)$ 
(see Fig. \ref{fig:surface}):
\begin{equation}
\omega = -2 [1 + \mu |A|^{2}]\cos k - 2\nu |A|^{2}\;.
\label{dispersion}
\end{equation}
Note that due to the nonlinear character of the eq.
(\ref{Salerno}), the frequency $\omega$ depends on both wave number
$k$ and amplitude $|A|$ of the plane wave.

One can easily determine those plane waves that are ($p$, $q$)
resonant with respect to $\omega_b$: the eq.
(\ref{resonance}) imposes the following condition on $\omega$ and
$k$
\begin{equation}
\frac{\omega}{\omega_b} = \frac{1}{q}\left(\frac{p}{2\pi} k -m
\right)\;, \label{resonant}
\end{equation}
where $m$ is any arbitrary integer. These planes in the 3-d space
($\omega$, $|A|$, $k$) intersect the dispersion relation surface
at (in general) several one-parameter families (branches)
$k_j(|A|)$, in the first Brillouin zone ($-\pi \le k \le \pi$).

If we are not interested in unreasonably large (and not interesting)
amplitude values $|A|$ of the plane waves, the number of branches
is finite: one can see that for fixed values of all the parameters
($p$, $q$, $\omega_b$, $\nu$, $\mu$), there is a finite number of
branches in the limit $|A| \rightarrow 0$; there is also a well
defined (parameter dependent) treshold value of the amplitude at
which a pair of new branches (tangent bifurcation) appear ({\em
i.e.} these plane waves can only resonate with $\omega_b$ for
amplitudes above some threshold value).

Thus, by a suitable bounding of the amplitude, for each couple
($p$, $q$) one finds a finite number, $s$, of branches of ($p$, $q$)
resonant plane waves. (Note also that this number diverges when
$p/q$ tends to an irrational).

A different, and highly nontrivial, example of ($p$, $q$)
resonant solutions is provided by the solitary waves
(\ref{A-Lbreather}) of the A-L lattice. From eq.
(\ref{omegav}) it is clear that the choice $2\pi v_b/\omega_b =
p/q$ selects a ($p$, $q$) resonant solitary wave with respect to
the frequency $\omega_b$. The set of velocity values of resonant
A-L breathers is dense and any A-L moving breather is a limit of
some sequence of resonant ones. Note also that immobile breathers
are (0, 1) resonant with respect to the frequency $\omega_b$.

In the integrable limit, the plane waves and the A-L breathers are
both exact independent solutions. Integrability makes possible
that the initial localization of energy is maintained with time
evolution, without decaying away by exciting radiation. It is a
well established result that (even far away from this integrable
limit) immobile discrete breathers remain exact solutions of
the lattice dynamics. Our concern in the next sections is the
question of moving discrete breathers away from integrability in
eq. (\ref{Salerno}). In order to study them, we will focus
on ($p$, $q$) resonant solutions. The motivating of
this restriction comes from its accessibility to numerics. First
we will motivate the numerical (Newton) method that allow us to
study these solutions with an adequately high precision.

\subsection{Newton continuation}
\label{Newton}

A well-known numerical procedure to obtain exact periodic
solutions of nonlinear lattices is the Newton continuation
\cite{Marin, CretegnyAubry,Cretegny,MackayAubry}. The different practical
implementations of this procedure work very successfully when, for
example, one obtains numerically exact immobile discrete breathers of
eq. (\ref{Salerno}), from the uncoupled limit $\mu =0$ and
$C=0$, where exact periodic discrete breathers are trivially
constructed.

The iteration of the Newton operator ${\cal T}$ converges rapidly
to its fixed point ({\em i.e.} the solution to be computed)
provided the starting point, $\hat\Phi^{0}$, is close enough, 
and the solution of the following system of linear equations 
is a well-posed problem:
\begin{equation}
( D{\cal T} - 1 ) (\Phi^n - \Phi^{n+1}) = [{\cal T} - {\cal I}]
\Phi^n \label{Newton_iteration}\;,
\end{equation}
where $D{\cal T}$ is the jacobian matrix of the Newton operator,
and $\Phi^n$ (the $n$-th iteration solution of
(\ref{Newton_iteration})) converges quadratically to the fixed point 
solution. By adiabatic change of a model parameter, one constructs a
uniquely continued exact fixed point solution for each parameter
value, using each time, as starting point of the Newton iteration,
the solution previously computed.

The matrix $( D{\cal T} - 1 )$ must be invertible, in order to
uniquely compute $\Phi^{n+1}$. Degeneracies associated with the
$+1$ eigenvalues of $D{\cal T}$, if any, have to be removed in
order to obtain a unique fixed point solution. When continuing
immobile (periodic) discrete breathers of eq.
(\ref{Salerno}), a convenient prescription is commonly used,
namely to restrict the operator action to the subspace of
time-reversible solutions \cite{Marin,Cretegny}. This provides a
practical way of removing degeneracies, allowing unique
continuation of immobile discrete breathers.

For the continuation of ($p$, $q$) resonant solutions (of which
periodic solutions are only the particular case $p=0$ and $q=1$),
one has to use ${\cal M}={\cal L}^{p} {\cal T}^{q}$ as the Newton
operator. One has also to deal with the degeneracies of ${\cal
M}$, and imposing time-reversibility could, in this case, be too
restrictive.

A well-known solution to the problem of removing degeneracies when
no clear restrictions are available, is provided by the so-called
{\em singular value decomposition} (SVD)
\cite{CretegnyAubry,Cretegny,Strang, NumRec} of the matrix $(
D{\cal L}^p{\cal T}^q - 1 )$ :
\begin{equation}
( D{\cal L}^p{\cal T}^q - 1 )= J = P V Q\;,
\end{equation}
where $P$, $V$ and $Q$ are $2N \times 2N$ square matrices. $P$ and
$Q$ are orthogonal matrices and $V$ is diagonal ($v_j
\delta_{ij}$) with possibly null (zero) elements, called singular
values, associated with the null space of $J$ (the subspace that is
mapped to zero $Jx=0$). The columns of $P$ whose same-numbered
elements $v_j$ are nonzero are an orthonormal set of basis
vectors that span the range of $J$ (the subspace reached by this
matrix). The rows of $Q$ whose same-numbered elements $v_j$ are
zero are an orthonormal basis for the null space of $J$. One can
numerically use this SVD decomposition, checking the (numerical)
vectors spanning the null space to identify degeneracies, and
using at iteration steps the pseudoinverse matrix
\begin{equation}
Q^* \hat{V}^{-1} P^*\;, \label{SVD}
\end{equation}
where $\hat{V}^{-1}$ is diagonal with elements $1/v_j$ for $v_j
\ne 0$ and $0$ for $v_j = 0$.

As a judicious test of our numerical codes, we have used both
procedures (reduction to time-reversible subspace and SVD
decomposition) to obtain immobile discrete breathers of the
Salerno model. Both agree, up to the highest possible accuracy,
from the uncoupled limit (one-site and two-site
breathers) up to the A-L limit (and viceversa).

This test serves also to provide further confirmation of an
important and well-known theoretical result. At the integrable
A-L lattice, one-site and two-site immobile breathers
are but two particular choices of the continuous one-parameter
($x_0$, the localization center) family of immobile solitary
waves, {\em i.e.} constant $x_0(t)=n$ or $n+1/2$ respectively, in
eq. (\ref{A-Lbreather}). The well-known result, confirmed by
our numerics, is that away from the A-L limit {\em only} these
(one-site and two-site) immobile discrete breathers persist under
adiabatic continuation. No immobile breather centered in between
exists. For positive values of the parameter $\nu$, the one-site
immobile one has a lower value of energy ${\cal H}$, and it is a 
linearly stable solution, while the energy of the two-site breather 
is higher and it is linearly unstable. The relative situation is
reversed for negative values of $\nu$. This result can be
interpreted as the emergence of a (Peierls-Nabarro) potential
function of the breather center $x_0$, which destroys the
continuous degeneracy of immobile breathers, leaving only two of
them per lattice unit, namely those localized at maxima and minima
of the Peierls potential. This interpretation, which is captured 
in the theoretical framework of collective variable
approaches, turns out to play a central role in building up the
physical interpretation of the numerical results on mobile
discrete Schr\"odinger breathers, below in section \ref{sec:Mobile}.

The numerical integration of the equations was performed using a
$4^{th}$ order Runge-Kutta scheme with time step $T_{b}/500$. 
The convergence criterion for the fixed point solution is
that the value of $\sum_{j}|([{\cal T} - {\cal I}]\Phi^{n+1})_{j}|$ is less than 
$N\cdot10^{-16}$ (where N is the size
of the lattice). The typical size of the lattices was taken between
$100$ and $200$ sites depending on the characteristics of the solution
considered, as we will explain in section \ref{sec:Mobile}.

\subsection{Floquet stability analysis}
\label{Floquet.Ana}

A very useful outcome of the numerical Newton method of computing
solutions of eq. (\ref{Salerno}) is the jacobian matrix of
the Newton operator, usually called the Floquet matrix $F$. This
matrix is the linear operator associated with the linear stability
problem \cite{Marinetal} of the fixed point solution.

Indeed, the jacobian $F$ of the Newton operator ${\cal M}$
\begin{equation}
F =  D{\cal M} \label{Floquetmatrix}
\end{equation}
maps vectors in the tangent space of the solution (small initial
perturbations $\vec{\epsilon}(0)$ of the fixed point solution)
into their $T_{{\cal M}}$-evolved vectors, {\em i.e.}
$\vec{\epsilon}(T_{{\cal M}})$, after a period of ${\cal M}$. 
That is:
\begin{equation}
\vec{\epsilon}(T_{{\cal M}})={\cal F}\vec{\epsilon}(0)\;,
\label{Floquet_Matrix}
\end{equation}

The Floquet matrix of a Hamiltonian system is real and symplectic,
so the Floquet eigenvalues $\lambda$ come in quadruplets,
$\lambda, 1/\lambda, \bar{\lambda}, 1/\bar{\lambda}$. The
necessary condition for the stability of the solution is that all
the eigenvalues lie on the unit circle of the complex plane,
$|\lambda|=1$.

To illustrate the Floquet analysis of ($p$, $q$) resonant
solutions of the NLS lattice (\ref{Salerno}),
we now obtain the Floquet spectrum of modulational instabilities
of a ($p$, $q$) resonant plane wave,
\begin{equation}
\Phi_n(t)=A\exp{{\mbox i}(kn-\omega t)}\;. \label{planewave}
\end{equation}
(The modulational instabilities of plane wave solutions of
nonlinear lattices have been analyzed in
\cite{KivsharPeyrard,KivsharSalerno}).

One has to investigate the evolution of small perturbations, in
both amplitude and phase, of the plane wave
\begin{equation}
\Phi_n(t)=(A+I_n)\exp{{\mbox i}(kn-\omega t+\varphi_{n})}\;,
\label{planewave_pert}
\end{equation}
where we assume that the perturbation parameters are small
compared with those of the plane wave solution. Introducing
expression (\ref{planewave_pert}) in (\ref{Salerno}) and
considering the following form for the perturbations $\lbrace I_n,
\varphi_{n}\rbrace$:
\begin{eqnarray}
I_n(t)&=&I\exp{{\mbox i}(Qn-\Omega t)} \nonumber
\\
\varphi_n(t)&=&\varphi\exp{{\mbox i}(Qn-\Omega t)}
\label{planewave_pert1}
\end{eqnarray}
we obtain the dispersion relation for the perturbation parameter
$\Omega$:
\begin{eqnarray}
\nonumber \lbrack \Omega -2(1+\mu A^{2})\sin{k}\sin{Q}\rbrack^2&=&
16(1+\mu A^2)\times
\\
\sin^2{Q/2}\cos{k} \lbrack(1&+&\mu A^2)\sin^2{Q/2}\cos{k}
\nonumber
\\
&-&\mu A^2\cos{k} -\nu A^2 \rbrack, \label{disp_pert}
\end{eqnarray}
From the above expression one derives the values of $\Omega
(A,Q,k;\nu,\mu)$ for the modulational perturbations. When the
parameter $\Omega$ has a nonzero imaginary part, {\em i.e.} the
right-hand side of (\ref{disp_pert}) is negative, the plane wave
($A, k$) becomes unstable under the corresponding modulational
($Q$) perturbation, whose amplitude will grow exponentially fast
in the linear regime (tangent space).

Modulational perturbations (\ref{planewave_pert1}) correspond to
eigenvectors $\lbrace I_n, \varphi_{n} \rbrace$ of the Floquet
matrix:
\begin{eqnarray}
I_{n}(t+T_{\cal M})=\exp(-{\mbox i}\Omega T_{\cal M})I_{n}(t)
\\
\varphi_{n}(t+T_{\cal M})=\exp(-{\mbox i}\Omega T_{\cal
M})\varphi_{n}(t) \label{evol_MI}
\end{eqnarray}
with associated Floquet eigenvalues $\exp(-{\mbox i}\Omega
T_{{\cal M}})$. The real part of $\Omega$ gives the angle in the
complex plane,
\begin{equation}
\theta_{Floq}=-\Re(\Omega)T_{{\cal M}}\;, \label{Floquet_angle}
\end{equation}
while the imaginary part $\Im(\Omega)$ gives the modulus of the
Floquet eigenvalue,
\begin{equation}
|\lambda|=\exp(\Im(\Omega) T_{{\cal M}})\;,
\end{equation}
thus providing the information about the linear stability of the
solution.

The distribution of angles and moduli in the Floquet spectrum of
the modulational instability can be obtained from eq.
(\ref{disp_pert}) by taking the real and imaginary parts of
$\Omega$:
\begin{eqnarray}
\Re(\Omega)&=&2(1+\mu A^{2})\sin{k}\sin{Q}
\label{Re}
\\
\Im(\Omega)^2&=&-16(1+\mu A^2)\sin^2Q/2\cos{k}\times \nonumber
\\
&\times& \lbrack(1+\mu A^2)\sin^2{Q/2}\cos{k} \nonumber
\\
&-&\mu A^2 \cos{k} -\nu A^2 \rbrack\;.
\label{Im} 
\end{eqnarray}

\begin{figure}[!tbh]
\begin{tabular}{cc}
\centerline{\resizebox{8cm}{!}{%
\includegraphics[angle=-90]{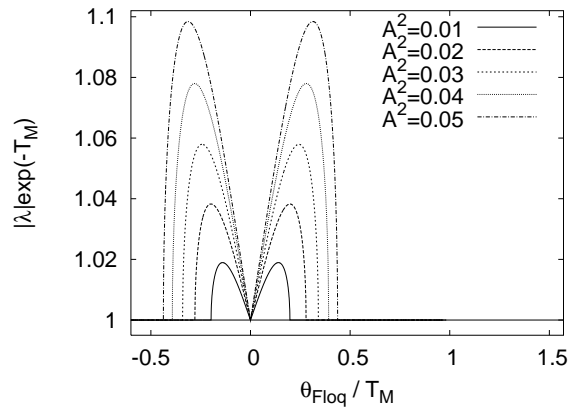}
}}
\end{tabular}
\caption{Plot of the modulus of the unstable Floquet eigenvalues 
$|\lambda|$ (corresponding to the positive values of $\Im(\Omega)$ in
eqs. (\ref{Re}) and (\ref{Im})),
versus the Floquet angle, $\theta_{Floq}$. Both quantities are
conveniently normalized to the period of the map $T_{\cal M}$. The
amplitude of the excursion of $|\lambda|$  and the range of values of
$\theta_{Floq}$ for which $|\lambda|>1$ grow as the amplitude $A$ of
the plane wave is increased. The parameters in eq. (\ref{Salerno})
are $\mu=\nu=0.5$ and the wave number of the plane wave is $k=0.5$.} 
\label{fig:MIfloq}
\end{figure}

In Fig. \ref{fig:MIfloq}
we represent the modulus of the unstable eigenvalues as a function
of the Floquet angle for the spectrum of a ($p$, $q$) resonant
plane wave, taken as an example to visualize the non-point-like
character of the instability in the Floquet spectrum in the
thermodynamic limit. Note that there is no plane wave harmonic
instability ($\theta_{Floq} = 0$) due to this mechanism of
modulational instabilities.

\begin{figure}[!tbh]
\begin{tabular}{cc}
\centerline{\resizebox{8cm}{!}{%
\includegraphics[angle=-90]{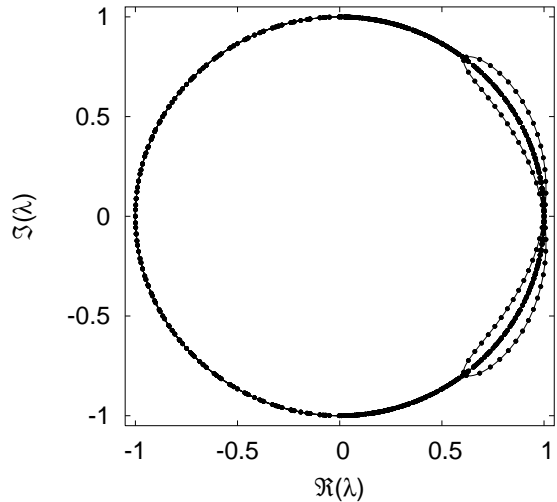}
}}
\end{tabular}
\caption{Plot of the Floquet spectra of a plane wave with modulational
instability (circles) and the theoretical prediction (lines) for the
distribution of the Floquet eigenvalues in the complex plane given by
eqs. (\ref{Re}) and (\ref{Im}). The amplitude and wave number of the plane
wave are $A=0.1$ and $k=0.1\cdot2\pi$; the nonintegrable parameter
value is $\nu=0.1$ and the lattice size is of $400$ sites.} 
\label{fig:miPW}
\end{figure}

A numerical computation of the Floquet spectrum of a plane wave
(with arbitrary wave number) of a lattice of $N=400$ sites,
with periodic boundary conditions is shown in the complex plane
representation of Fig. \ref{fig:miPW}. The instability globes,
at angles symmetrically placed around zero in this figure, nicely
fit the theoretical (thermodynamic limit) values obtained from
eqs. (\ref{Re}) and (\ref{Im}).

\section{Mobile discrete Schrodinger breathers}
\label{sec:Mobile}

In this section, we show the numerical results on mobile discrete
Schr\"odinger breathers in the NLS lattice (\ref{Salerno}). These
numerics are computed using the tools explained in the previous
section. The Newton fixed point continuation requires a good
initial guess (meaning that the starting initial conditions have
to be in a small neighborhood of the fixed point). Very close to
$\nu = 0$, the A-L solitary traveling waves (exact solutions at
$\nu=0$) provide good starting points. After convergence to the
fixed point, we increase adiabatically the value of the parameter
($\Delta \nu = 10^{-3}$), and start iteration from the previous
fixed point.

An important step in the numerical method used here, is obtaining
a basis for the subspace of (tangent space) vectors with Floquet
eigenvalue $+1$. These are associated to those degeneracies
(symmetries) that one has to eliminate in order to regularize the
linear system at each (Newton) iteration step when numerically
converging to the fixed point solution.

Away from the A-L limit, it is known (as reported {\em
e.g.} in \cite{Dmitriev}) that only two conserved quantities
remain generically as dynamical invariants, the Hamiltonian (\ref{Salerno_Ham})
and the norm (\ref{norm}). They are respectively associated to the
continuous time translation and gauge (global phase rotation)
invariance. Using the notation $u_i=\Re (\Phi_i)$ and 
$v_i=\Im (\Phi_i)$, one easily obtains that 
($\delta u_i(t)=\dot{u}_i(t)$, $\delta v_i(t)=\dot{v}_i(t)$) 
is the perturbation associated with time translational invariance, 
while ($\delta u_i(t)=v_i(t)$, $\delta v_i(t)=-u_i(t)$) is the 
one with gauge invariance. These are, consequently, Floquet 
eigenvectors with associated eigenvalue $+1$, and we can easily 
check that they coincide with the (two) basis vectors provided 
generically ({\em i.e.} except at special bifurcation values of 
the parameter, see below in \ref{Flo.An}) by the numerical Singular 
Value Decomposition (\ref{SVD}) explained in the previous section.

In subsection \ref{Structure} we summarize our findings on the
generic structure of mobile Schr\"odinger discrete breathers. For
this, as explained earlier, we have explored particular values for
the integers ($p$, $q$) and performed continuation of ($p$, $q$)
resonant A-L traveling waves. The variation of the
main structural characteristics of the fixed points along the
continuation parameter $\nu$ is examined in detail in
\ref{Background.Amplitude}, for both signs of this parameter. Of
particular interest are the observed drastic changes in the
structure for $\nu \simeq -0.3$ and $\nu \simeq -0.39$. 
Then, in \ref{Flo.An}, we show the main conclusions on the 
stability analysis of the mobile Schr\"odinger discrete 
breathers, in a sector of the breather parameter space.

\subsection{The structure of the solution}
\label{Structure}

\begin{figure}[!tbh]
\begin{tabular}{cc}
\centerline{{\bf (a)}
\resizebox{6.5cm}{!}{%
\includegraphics[angle=-90]{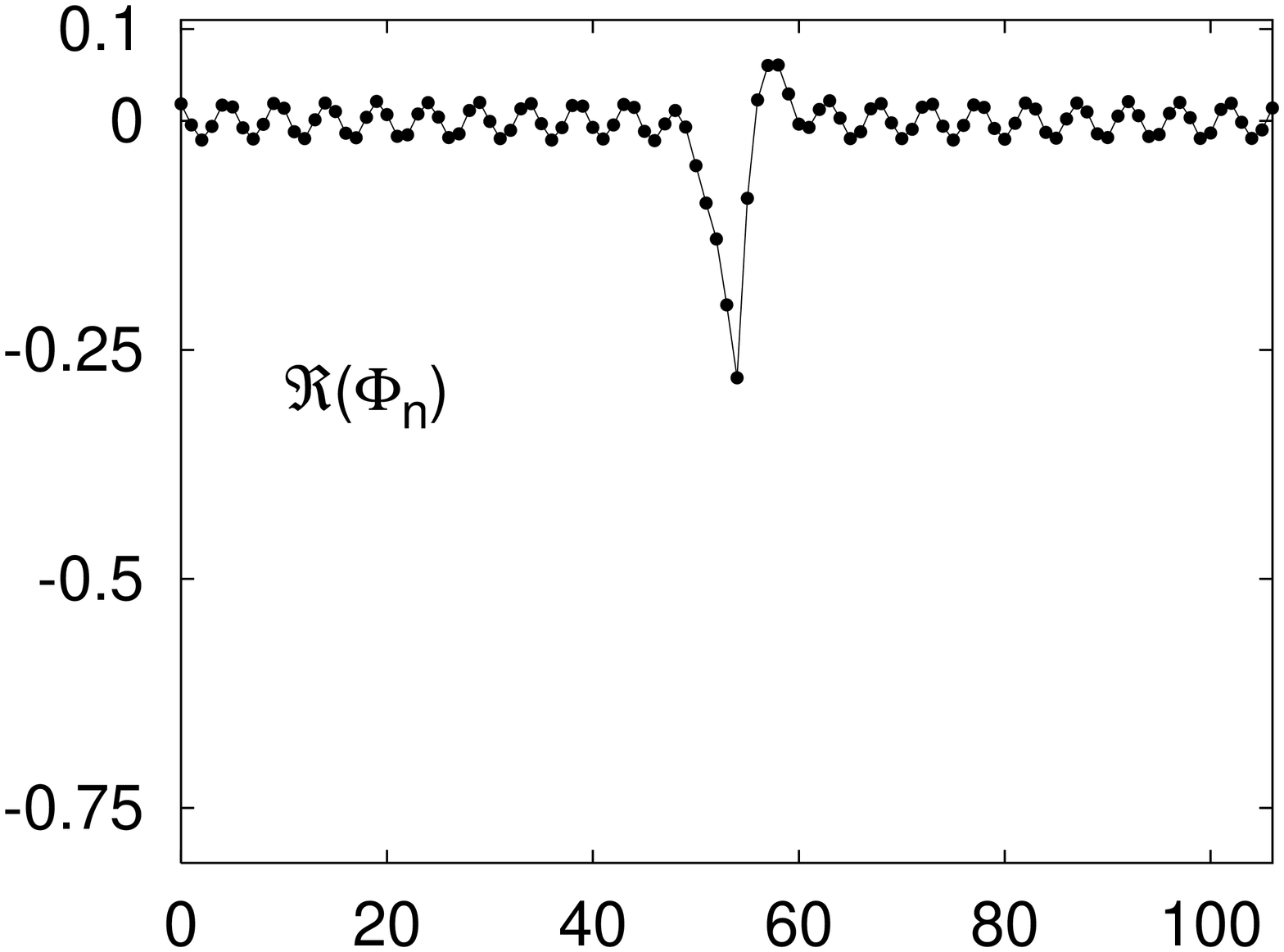}
}
{\bf (b)}
\resizebox{6.5cm}{!}{%
\includegraphics[angle=-90]{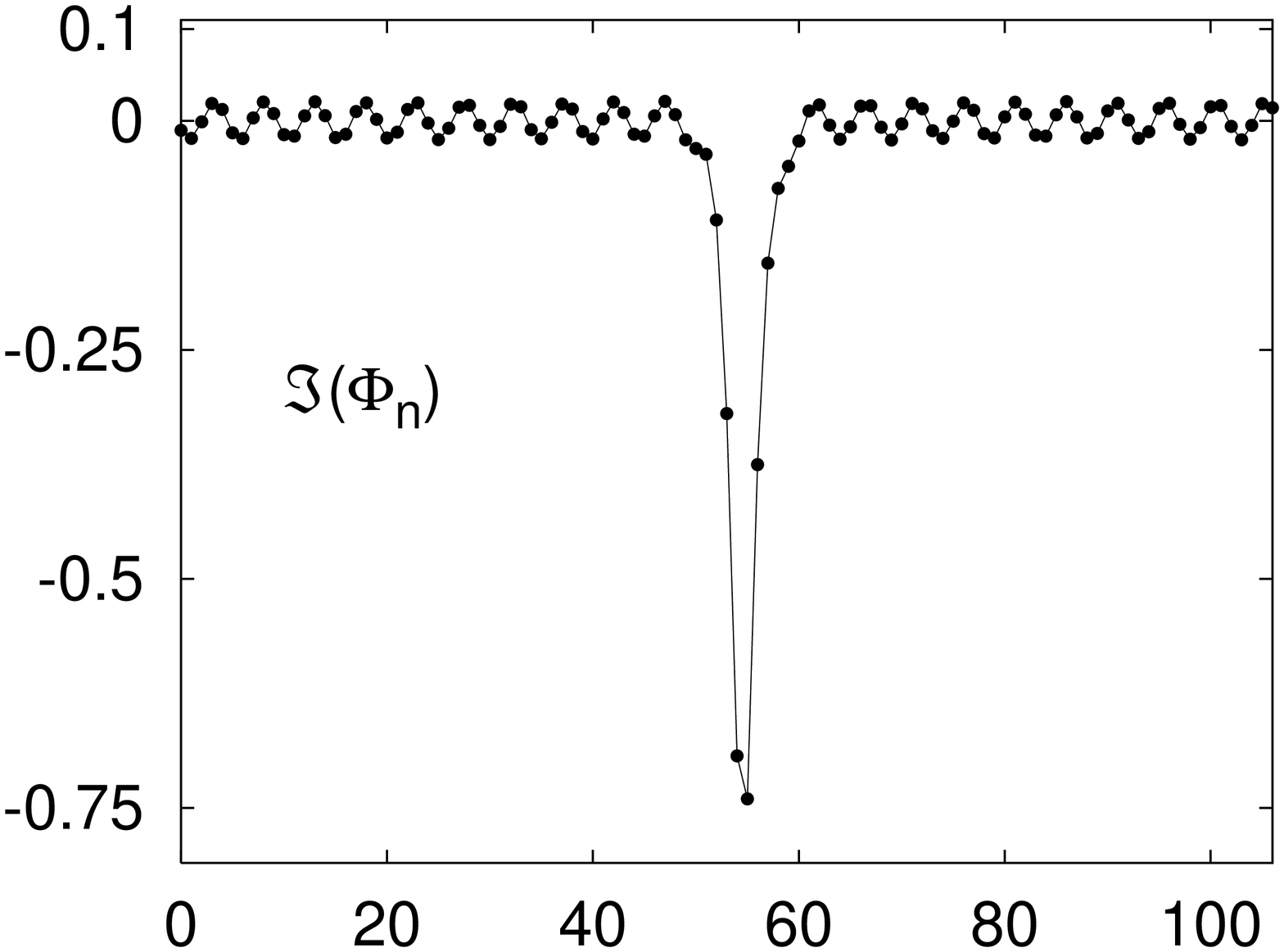}
}}
\\
\centerline{{\bf (c)}
\resizebox{6.5cm}{!}{%
\includegraphics[angle=-90]{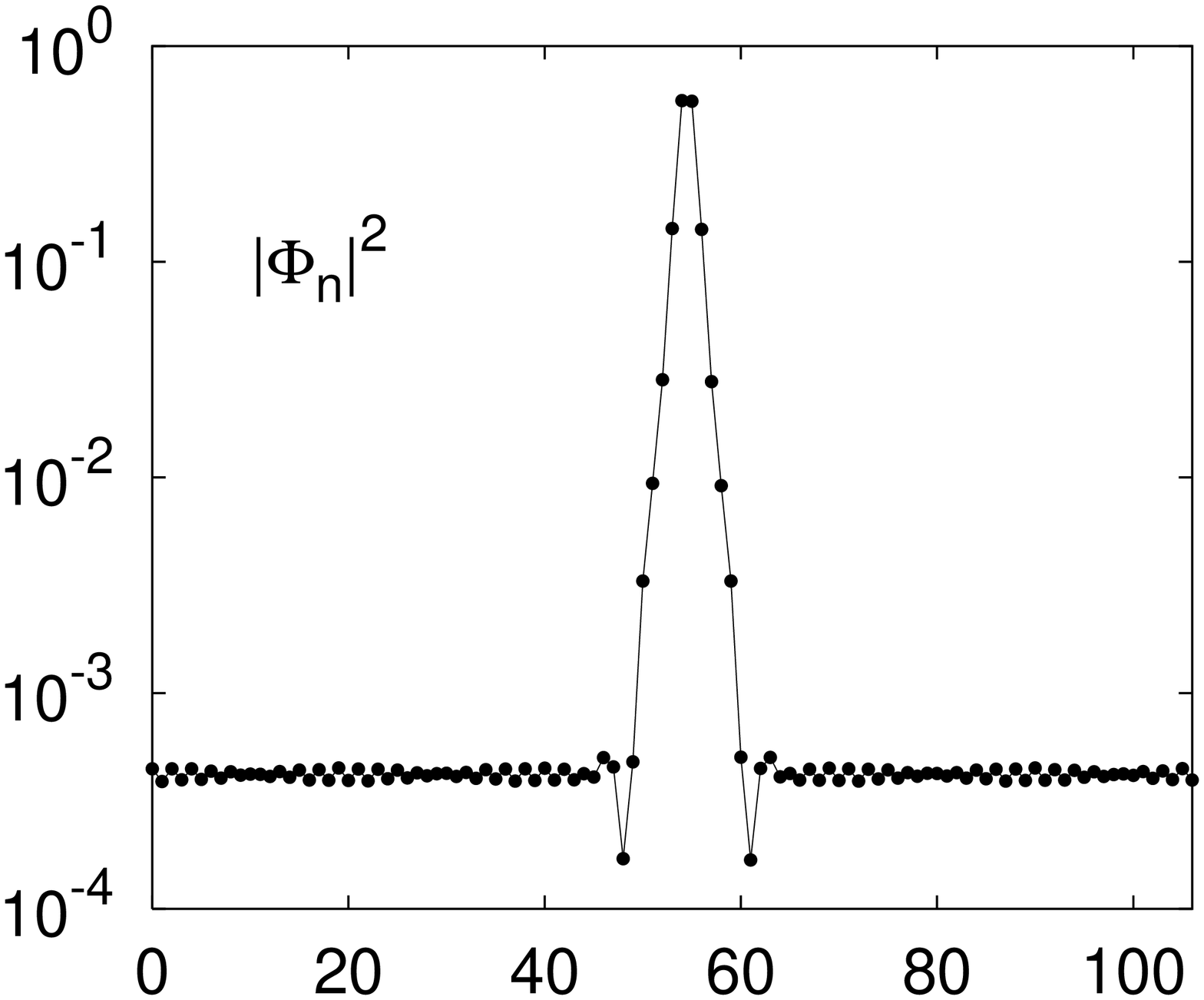}
}
{\bf (d)}
\resizebox{6.5cm}{!}{%
\includegraphics[angle=-90]{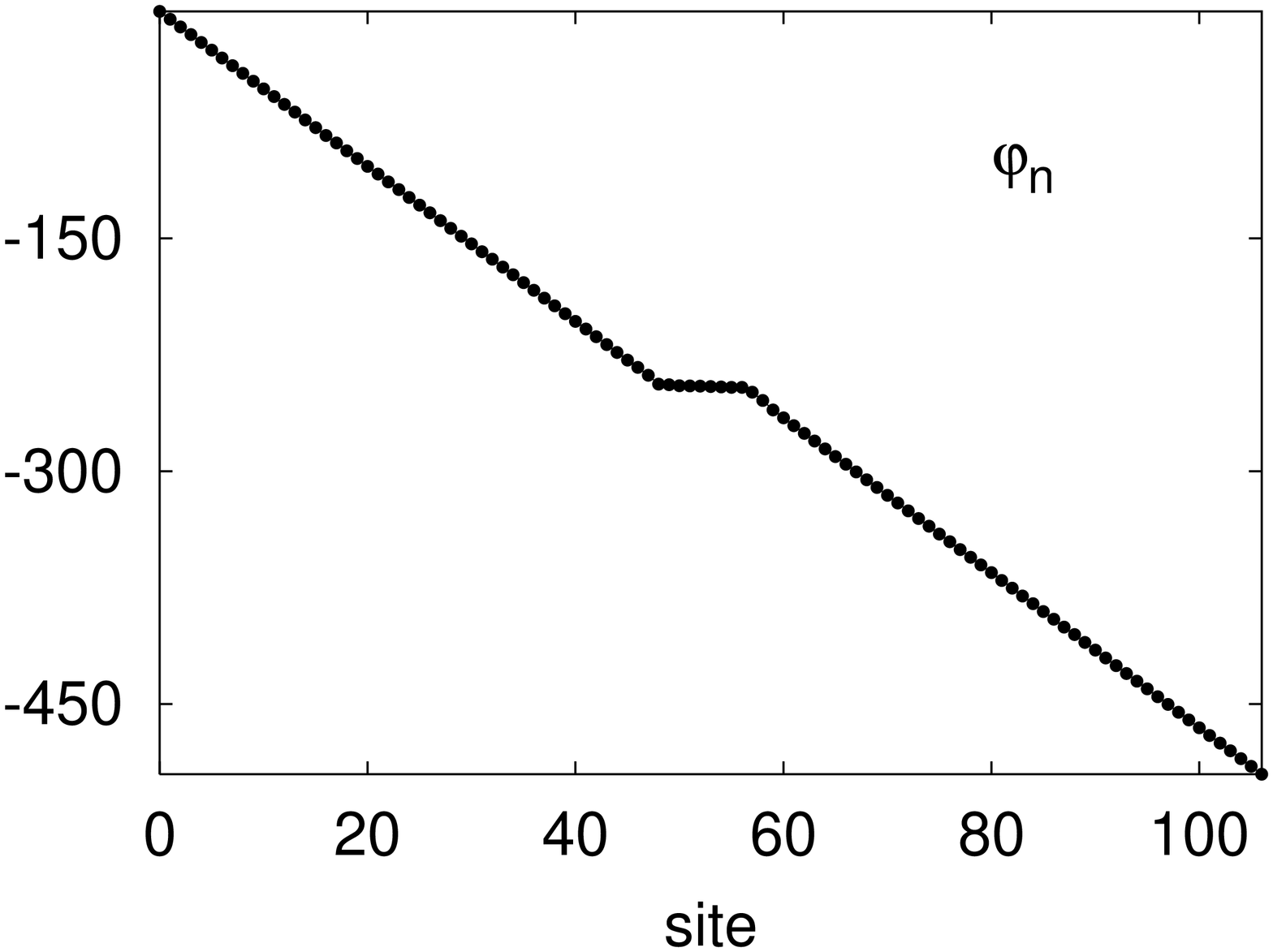}
}}
\end{tabular}
\caption{ 
Instantaneous profile of a $(1,1)$ resonant breather with
$\omega_{b}=2.678$ and $v_{b}=0.426$; the nonintegrable parameter 
is $\nu=1.0$ (standard DNLS equation). {\bf (a)} Real part, {\bf (b)} 
imaginary part, {\bf (c)} modulus and {\bf (d)} phase. 
The resonant condition for the harmonic composition of the 
background gives the contribution of three plane waves. The existence 
of these plane waves is revealed by the modulation of the extended 
tail in the modulus profile {\bf (c)}.} 
\label{fig:profile}
\end{figure}

In Fig. \ref{fig:profile} we plot the spatial profile of a ($1$, $1$) 
mobile Schr\"odinger discrete breather for nonintegrability parameter
value $\nu = 1.0$, and $\omega_b = 2.678$.

A quick inspection of this figure provides a first glance of the
general structure of the computed ($p$, $q$) resonant solutions:
The fixed point $\hat{\Phi}$ is the superposition of an
(exponentially) localized oscillation (the {\em core}) moving on
top of an extended {\em background}.
\begin{equation}
\hat{\Phi} = \hat{\Phi}_{{\mbox{ core}}} + \hat{\Phi}_{{\mbox
{backg}}}\;. \label{core-bckg}
\end{equation}
In other terms, far away from the core localization site $n_0$,
the solution does not tend to the rest state $\hat{\Phi}_n = 0$, but to
an extended excited state of the lattice , {\em i.e.} for
$|n-n_0|\gg 1$
\begin{equation}
\hat{\Phi}_n(t) = (\hat{\Phi}_{{\mbox{backg}}})_n (t) \neq 0\;.
\end{equation}

One easily realizes (for example, consider a site very far from
$n_0$) that the background has to be itself ($p$, $q$) resonant.
This can be quickly checked in our numerics: Indeed, the power spectrum 
$S(\omega)=|\int_{-\infty}^{\infty}\Re [\hat{\Phi}_n(t)]
\exp(\mbox{i}\omega t)dt|^2$ at a site $n$ far from $n_0$ reveals
a finite number of $s$ peaks $\omega_j$, $j=0,..., s-1$; one can check
that each $\omega_j$ numerically fits to a branch of ($p$, $q$)
resonant plane waves (see eq. ({\ref{plane-wave})); this
provides a set of amplitudes $A_j$, and finally one confirms that
the superposition of the ($A_j$, $\omega_j$) plane waves fits the
numerical solution $\hat{\Phi}_n(t)$.

While immobile discrete breathers can be described as a sort of
homoclinic (and time periodic) connection on the rest state, the
mobile localized core instead {\em connects} a specific linear
superposition of low amplitude nonlinear plane waves. One could
say that the localized core needs for its motion to "surf over" a
specific extended state of radiation:
\begin{equation}
(\hat{\Phi}_{{\mbox{backg}}})_n (t)= \sum_{j=0}^{s-1} A_j \exp{{\mbox
i}(kn-\omega_j t)}\;. \label{bckg}
\end{equation}

We note that among the members of the ($s$-parameter) continuous
family of ($p$, $q$) resonant plane waves (see Section I), the
fixed point solution contains only a particular member ($A_j$,
$\omega_j$) from each branch (see Fig. \ref{fig.fft}.a). 
This selection varies smoothly with
the (adiabatic) continuation parameter $\nu$. In particular, the
amplitude modulus $|A_j|$ selected increases smoothly from its
zero value at the integrable limit ($\nu =0$), for both signs of
$\nu$.

\begin{figure}[!tbh]
\begin{tabular}{cc}
\centerline{{\bf (a)}
\resizebox{8.cm}{!}{%
\includegraphics[angle=-90]{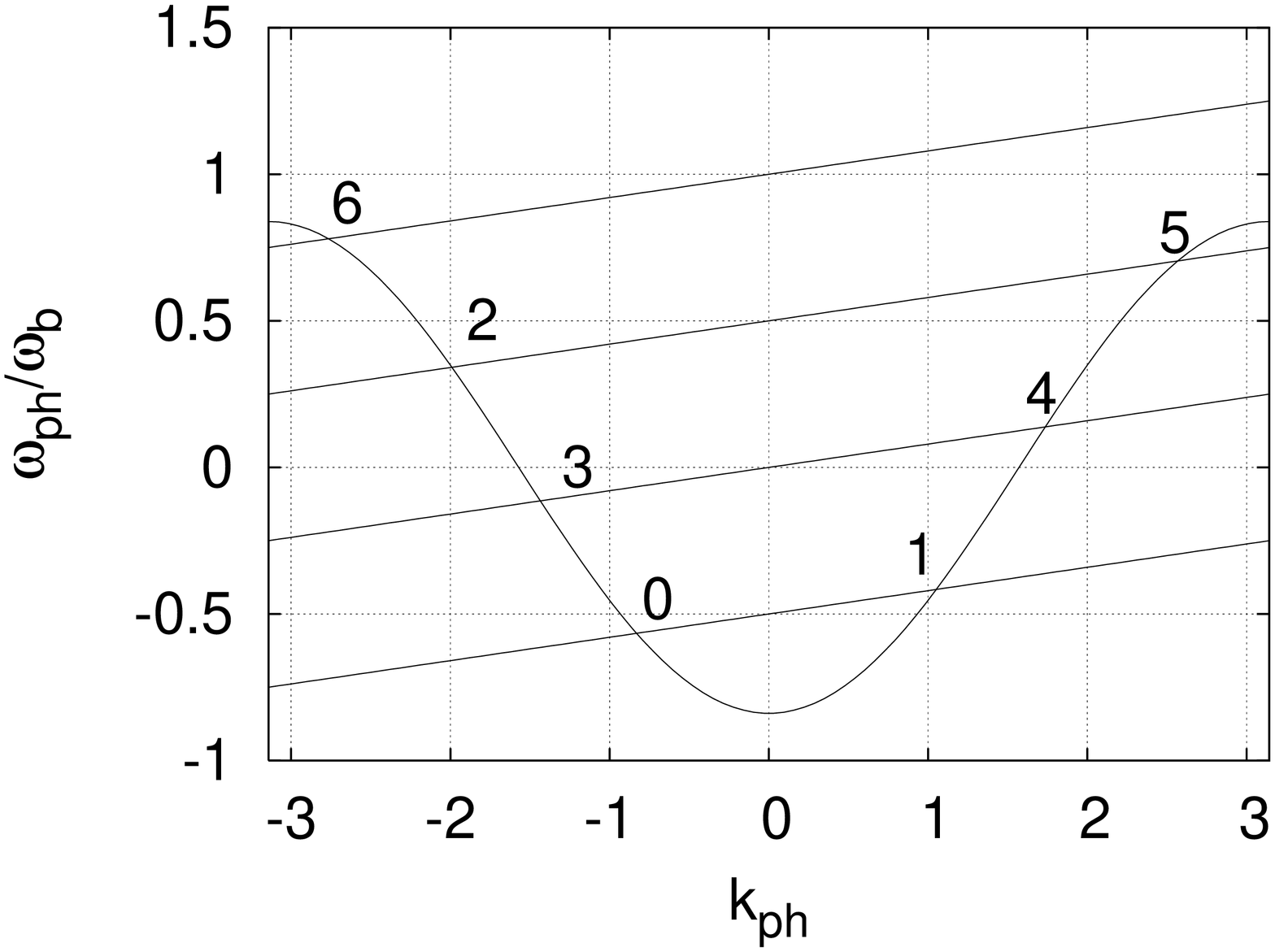}
}
{\bf (b)}
\resizebox{8.5cm}{!}{%
\includegraphics[angle=-90]{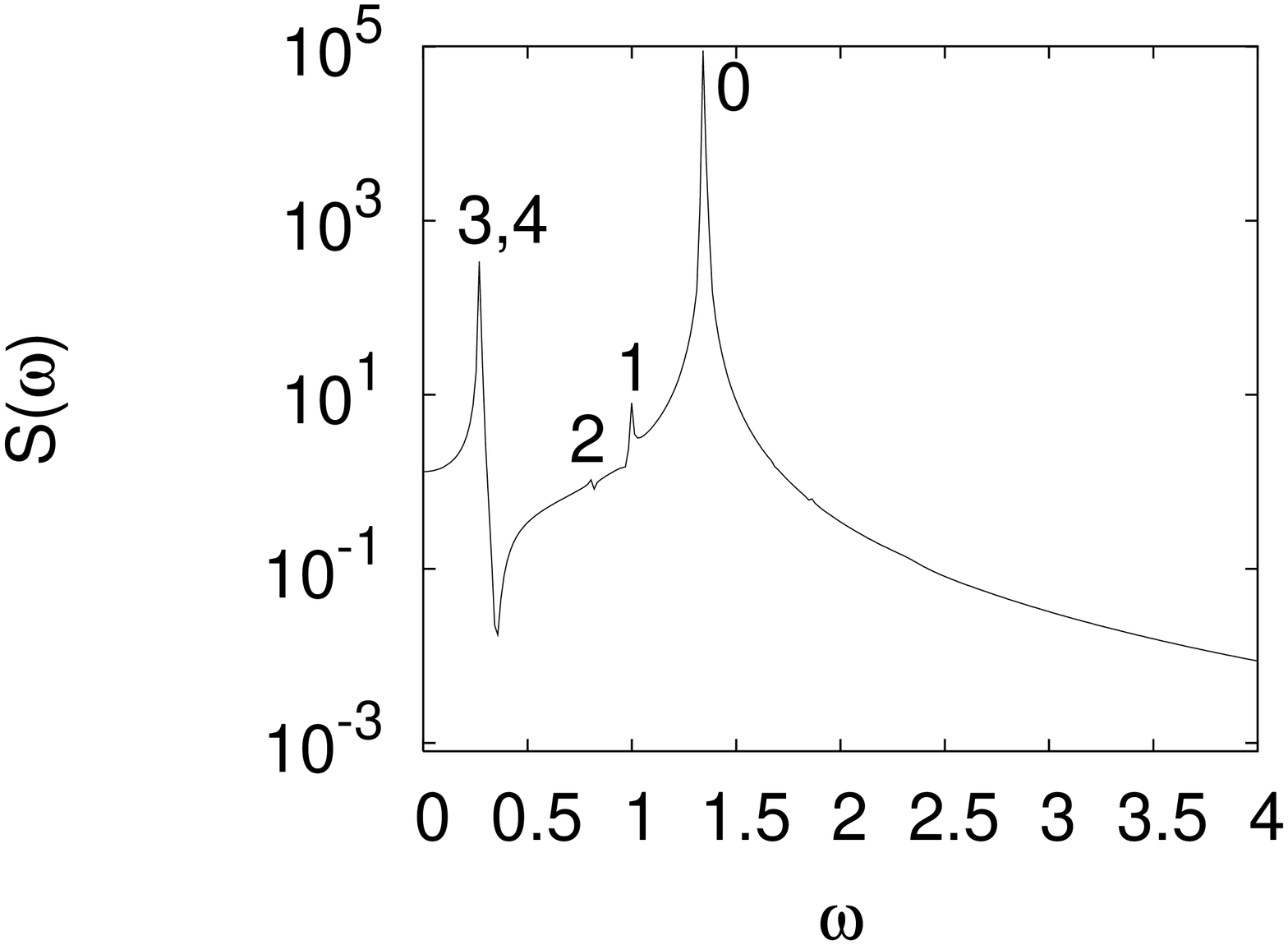}
s}}
\end{tabular}
\caption{{\bf (a)} Plot of the graphical solving of the resonant
condition (in the $A_{j}\rightarrow 0$ limit) for a $(1,2)$ resonant 
breather with $\omega_{b}=2.384$ and $v_{b}=0.189$. {\bf (b)} Power
Spectrum $S(\omega)$ of the background of this solution at $\nu=1.0$. 
From {\bf (a)} eq. (\ref{resonant}) gives the contribution of seven 
plane waves $(j=0,...,6)$ but only five $(j=0,...,4)$ of them are 
visible due to the difference  of orders of magitude between the 
amplitudes $|A_{j}|$. The agreement between the resonant condition 
equation (for the fitted value of $A_{j}$) and the frequencies
observed in $S(\omega)$ is up to machine accuracy.}
\label{fig:fft}
\end{figure}

If the bare core of a fixed point solution ({\em i.e.} after
substraction of the background) is taken as initial condition for
a direct numerical integration of the equations of motion, one
observes radiative losses, along with the corresponding changes in
shape, velocity, etc. of the localized moving core. The motion of
the bare localized core (not anymore a solution) excites extended
states of the lattice. Thus, regarding the exact fixed point
solution, one could say that radiative losses of the running core
are {\em exactly} cancelled out when the localized core runs, with
specific velocity, on top of the specific linear combination of
($A_j$, $\omega_j$) resonant plane waves (\ref{bckg}).

A complementary numerical observation is the following: Taking as
initial condition for a direct integration of the equations of
motion (\ref{Salerno}), a superposition of an immobile discrete
breather and the background of a ($p$, $q$) resonant mobile
breather, it evolves into a moving discrete breather, with
approximate velocity $v_b = (p\omega_b)/(2\pi q)$. One thus would
say that the background promotes breather translational motion
with adequate velocity. In the next section, a connection
between background characteristics and the particle perspective
({\em i.e.} the Peierls-Nabarro barrier of collective variable
theories), will be established, further illuminating the physical
description of discrete breather mobility.

Whatever physical perspective one may prefer, the numerical fact
is that the generic structure of the fixed point solution is given
by the superposition (\ref{core-bckg}). Not too far from $\nu
\simeq 0$, where the amplitudes $A_j$ of the fixed point
background have small values, one can carefully check that if the
bare core is given as a starting guess for Newton iteration, this
converges well to the exact complete solution (core $+$
background), by developing the specific selection of $A_j$
amplitudes. This confirms the robustness of the numerics.

Though previous observations of nondecaying tails of numerically
accurate mobile discrete breathers in Klein-Gordon lattices
\cite{CretegnyAubry} and/or (solitary) traveling waves
\cite{Duncan} in self-focusing equations had been
reported (see also the interesting discussions on this issue in
\cite{Kladko} and \cite{FlachZolotaryuk}), 
no systematic study of those tails is known to us. 
However we clearly see that they are an essential part 
of the exact solution. As argued in the introductory section, 
the translational motion of a discrete breather introduces 
a new time scale. In a nonintegrable context, this fact 
unavoidably implies resonances with plane wave band
spectra, and an exact self-sustained moving DB solution could only
exist on top of a developed resonant background. This seems to
have been (with a few exemptions) not fully appreciated in most of
current litterature on mobile breathers, where the backgound is
most often either ignored or deliberately suppressed.

A notable feature of the plane wave content of the
background $\hat{\Phi}_{{\mbox{backg}}}$ is that the amplitude
modulus $|A_j|$ in (\ref{bckg}) differ by orders of magnitude, {\em
i.e.} $|A_1| \gg |A_2| \gg |A_3|...$, so that only a few
frequencies are dominant for most practical purposes 
(see Fig. \ref{fig.fft}.b). In other words, the extended 
background associated to a spatially localized moving core is, 
in turn, strongly localized in the reciprocal ($k$-space) lattice. 
The possible relevance of this observation is further discussed below 
in the concluding section.

\subsection{The background amplitude}

\label{Background.Amplitude}

In order to characterize the specific features of the
nonintegrable motion of discrete Schr\"odinger breathers, we focus 
here on the (perhaps) most remarkable among those
features: the background amplitude of the uniquely continued fixed
point. How does it evolve along the continuation path in parameter
space?

For positive values of $\nu$ we have followed the line in
parameter space $\mu + \nu = 1$ (see equation (\ref{Salerno})), 
while for negative values, we took the path $\mu - \nu =
1$. We do not expect other paths to make important differences. 
As stated earlier, near $\nu\simeq 0$, the amplitude grows from 
its zero value (at the integrable limit) for both signs of 
this parameter, for it is a nonintegrable effect. However, 
for larger values of nonintegrability $|\nu|$ the background 
amplitude evolution shows some important differences
for the two signs of $\nu$.

In Fig. \ref{fig:amp_nu} we plot the background amplitude 
(modulus) of the ($1$, $1$) resonant fixed point, versus the 
continuation parameter $\nu$, for three different values of 
the breather frequency $\omega_b$. For $\nu > 0$, one observes 
that the amplitude steadily increases with $\nu$ before continuation 
stops ({\em i.e.} Newton iteration ceases to converge beyond a 
certain maximum $\nu$ value). Note that the amplitude grows faster 
for higher values of the frequency, and that the continuation stops
(correspondingly) at a smaller value of $\nu$. This may suggest
that the failure of fixed point continuation is related to a
somewhat excessive growth of the background amplitude, an issue
that will be discussed later.

For $\nu < 0$, after an initial growth the background amplitude
decreases down to almost negligible values around $\nu \simeq
-0.3$, then grows and again decreases close to zero at $\nu \simeq
-0.39$, and so on, in progressively narrower intervals with larger
peak amplitude, until continuation stops. Most noticeable is the
fact that the intervals neither depend on the breather frequency
$\omega_b$ nor on the breather velocity $v_{b}$. Why do background
amplitudes decay so dramatically at those regions in parameter
space? An important hint is presented in the next section, where 
the Floquet stability analysis of immobile discrete breathers 
will show a coincident situation of mirror-symmetry breaking 
(and its absence for positive $\nu$ values).

For other values of $p$ and $q$ that we have numerically
investigated, the same features of the background amplitude
variation as shown in Fig. \ref{fig:amp_nu} are qualitatively 
reproduced.

\begin{figure}[!tbh]
\begin{tabular}{cc}
\centerline{\resizebox{8.cm}{!}{%
\includegraphics[angle=-90]{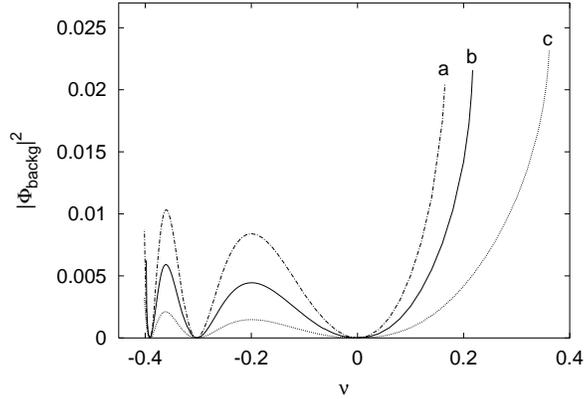}
}}
\end{tabular}
\caption{Background amplitude versus $\nu$ for three different 
$(1/1)$ resonant breathers with frequencies: (a) $\omega_{b}=5.65$, 
(b) $\omega_{b}=4.91$, (c) $\omega_{b}=4.34$. Note the two different
behaviours: for positive values of $\nu$ $|\Phi_{backg}|^2$ is a 
monotonous increasing function of $\nu$ while for the negative part it
shows smooth rises and falls.} 
\label{fig:amp_nu}
\end{figure}

\subsection{Floquet analysis}
\label{Flo.An}

On the basis of the general arguments given in
\cite{MarinAubry,Marinetal}, the Floquet spectra of immobile DB 
in the thermodynamic limit, $N \rightarrow \infty$, consists of 
two components: the (continuous) Floquet spectrum of the asymptotic
state of the solution (rest state), and a discrete part associated 
with spatially localized eigenvectors. The continuous part is composed
by small amplitude (linear) plane waves, the so-called
phonons. However, for mobile DB the asymptotic state of a ($p$, $q$) 
resonant fixed point solution is a superposition of plane waves, the
background $\hat{\Phi}_{{\mbox{backg}}}$. From this, one should expect the
Floquet spectrum of a ($p$, $q$) resonant DB being composed of 
two components: the discrete (spatially localized
eigenvectors) and a continuous part associated with the linear
stability of the background plane waves. The continuous part of the
Floquet spectrum should reflect the same results of the modulational
instabilty analysis of section \ref{Floquet.Ana}. In particular, this means that
any modulational instability a plane wave may suffer will be also an
instability of a fixed point solution whose background contains this
plane wave. In the future we will refer to to any instability of the
continuous part of the Floquet spectrum as {\em background
instability}. Any instability from the discrete part is a 
{\em core instability}. 

First we focus on {\em core instabilities}. For this we turn attention 
to the continuation of mobile ($p$, $q$) resonant breathers. 
Fig. \ref{fig:diagram} shows in the $\nu-\omega_b$ plane (dotted
line), the values $\nu_{max}(\omega_b)$ where the
numerical continuations stop due to nonconvergence of Newton 
iteration for $p=1$, $q=1$ and $\nu > 0$. 
As was remarked above, the continuation stop is associated with the
rapid increase of the background amplitude shown in
Fig. \ref{fig:fft}. Only low frequency breathers, for which the
background amplitude increases more slowly, can be numerically
continued all the way to the standard DNLS equation. The linear
stability analysis of ($p$, $q$) resonant breathers yields a well
defined region in the $\nu-\omega_b$ diagram where {\em core
instabilities} appear. There is an island inside the
continuation region of Fig. \ref{fig:diagram}, where the Floquet
spectra contain a real eigenvalue $\lambda > 1$. We observe the
evolution of this Floquet eigenvalue (and its complex conjugate)
as the parameter $\nu$ is increased in Fig. \ref{fig:uns_prof}.a,
for a (1, 1) breather of frequency $\omega_b = 2.678$. 
Here the angle ($\theta_{Floq}$) in the complex plane is plotted 
versus $\nu$. The interval of constant zero angle corresponds to the section
(constant $\omega_b$) of the instability island in Fig. \ref{fig:diagram}.

\begin{figure}[!tbh]
\begin{tabular}{cc}
\centerline{
\resizebox{8.cm}{!}{%
\includegraphics[angle=-0]{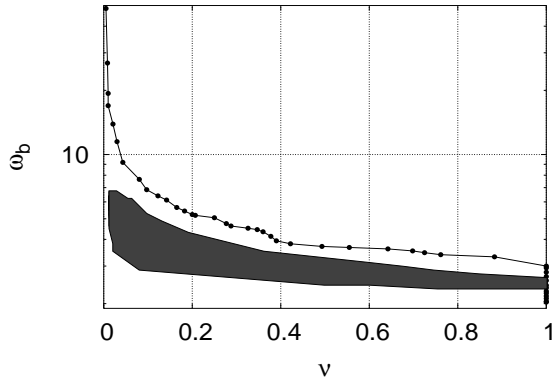}
}
}
\end{tabular}
\caption{Continuation diagram of $(1,1)$ resonant breathers as a functon of the
frequency $\omega_{b}$. The end of the numerical continuation,
$\nu_{max}(\omega_b)$, is represented by the line with dots. 
The region where mobile breathers
suffer from core instabilities is limited by the shaded area.} 
\label{fig:diagram}
\end{figure}

Along the whole continuation path, the profile of the corresponding 
unstable eigenvector is localized. An example of this profile inside the
instability island is shown in Figs. \ref{fig:uns_prof}.b and 
\ref{fig:uns_prof}.c, where one observes that the localized 
instability shows a decaying background along the direction opposite 
to the motion. The decay rate increases as the modulus of the 
eigenvalue grows and decreases again when $\lambda$ returns to the unit 
circle. On the other hand, the stable Floquet eigenvector 
associated with $1/\lambda$ shows a wing decaying along the mirror 
symmetric direction. The direct integration of the equation of 
motion reveals that the unstable solution experiences a pinning 
after a transient of regular motion with velocity $v_{b}=p/(qT_{b})$. 
After the solution pins at site $n$, its core center oscillates 
around this site. The trapping of the unstable MB could be interpreted 
as a result of the energy losses that the growth of the linearly
unstable perturbation induces on the solution.

Returning to the instability island shown in the diagram of
Fig. \ref{fig:diagram}, some final observations are worth
summarizing: (${\it i}$) there is a range of frequencies where
mobile breathers of the standard DNLS equation ($\nu=1$) suffer
from this instability; (${\it ii}$) very high frequency breathers
do not experience this instability (in the short range where they
can be continued); (${\it iii}$) very low frequency breathers are
stable all the way up to $\nu = 1$.

\begin{figure}[!tbh]
\begin{tabular}{cc}
\centerline{
{\bf (a)}
\resizebox{8cm}{!}{%
\includegraphics[angle=-90]{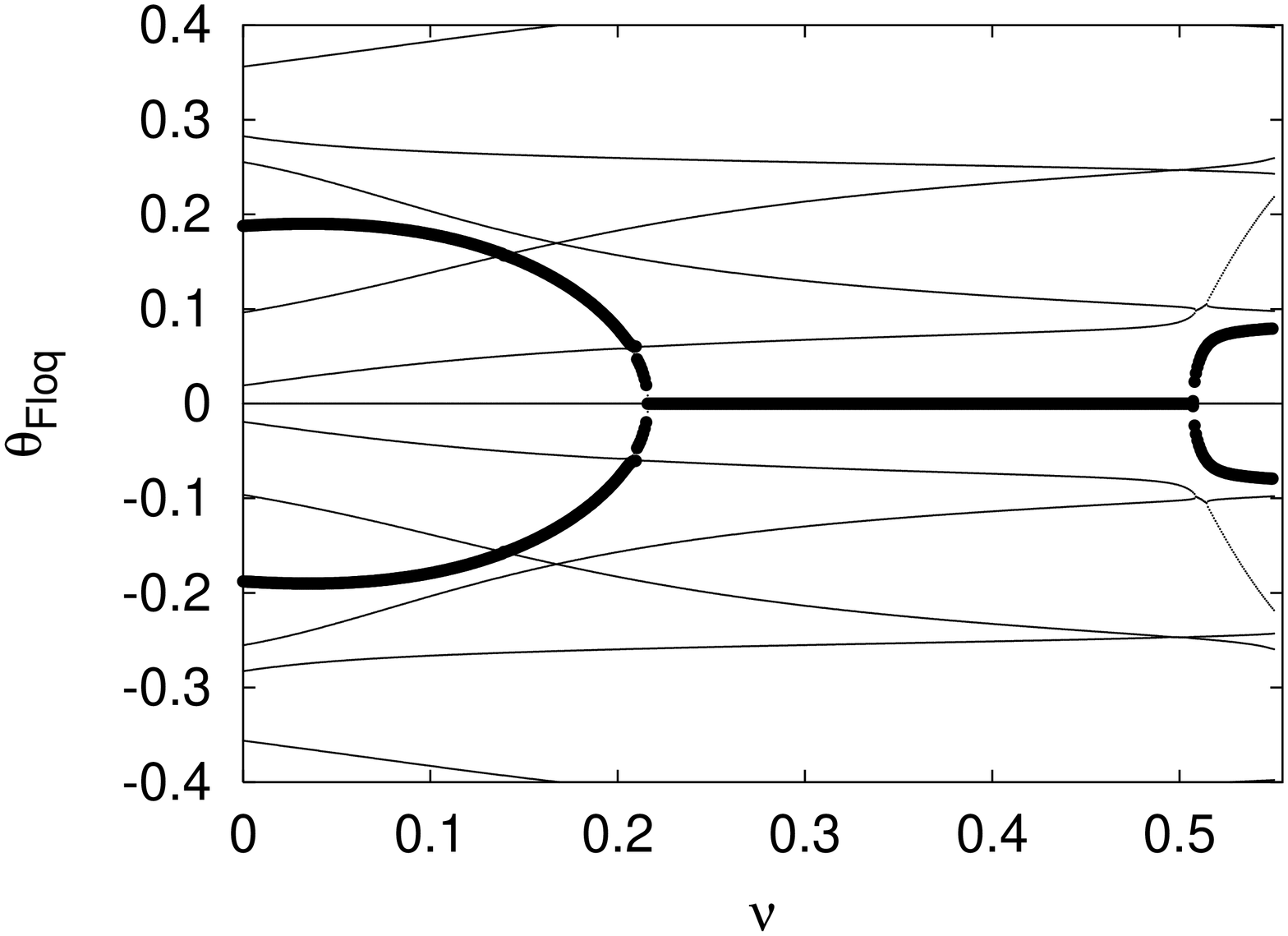}
}
}
\\
\centerline{
{\bf (b)}
\resizebox{8cm}{!}{%
\includegraphics[angle=-90]{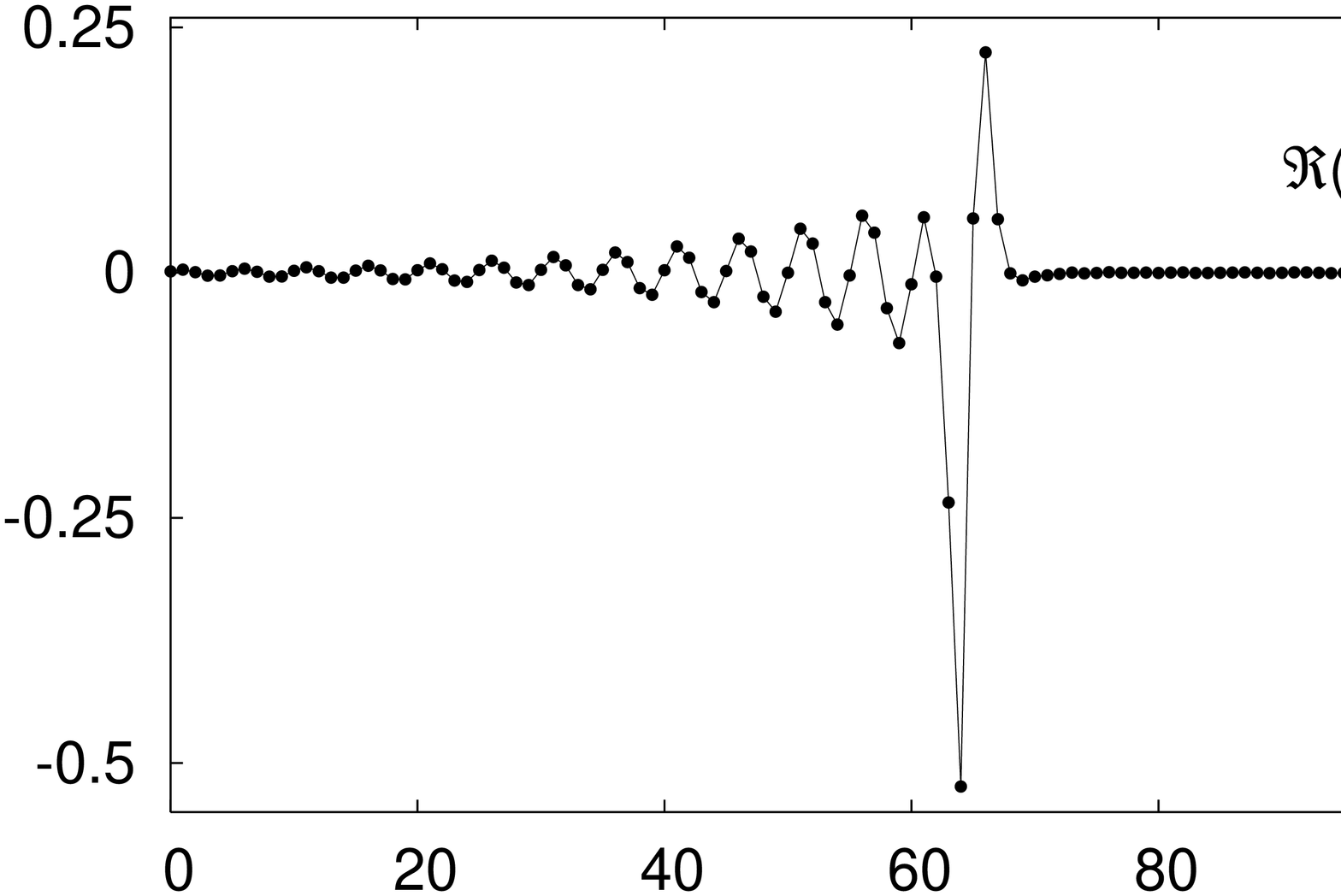}
}}
\\
\centerline{
{\bf (c)}
\resizebox{8cm}{!}{%
\includegraphics[angle=-90]{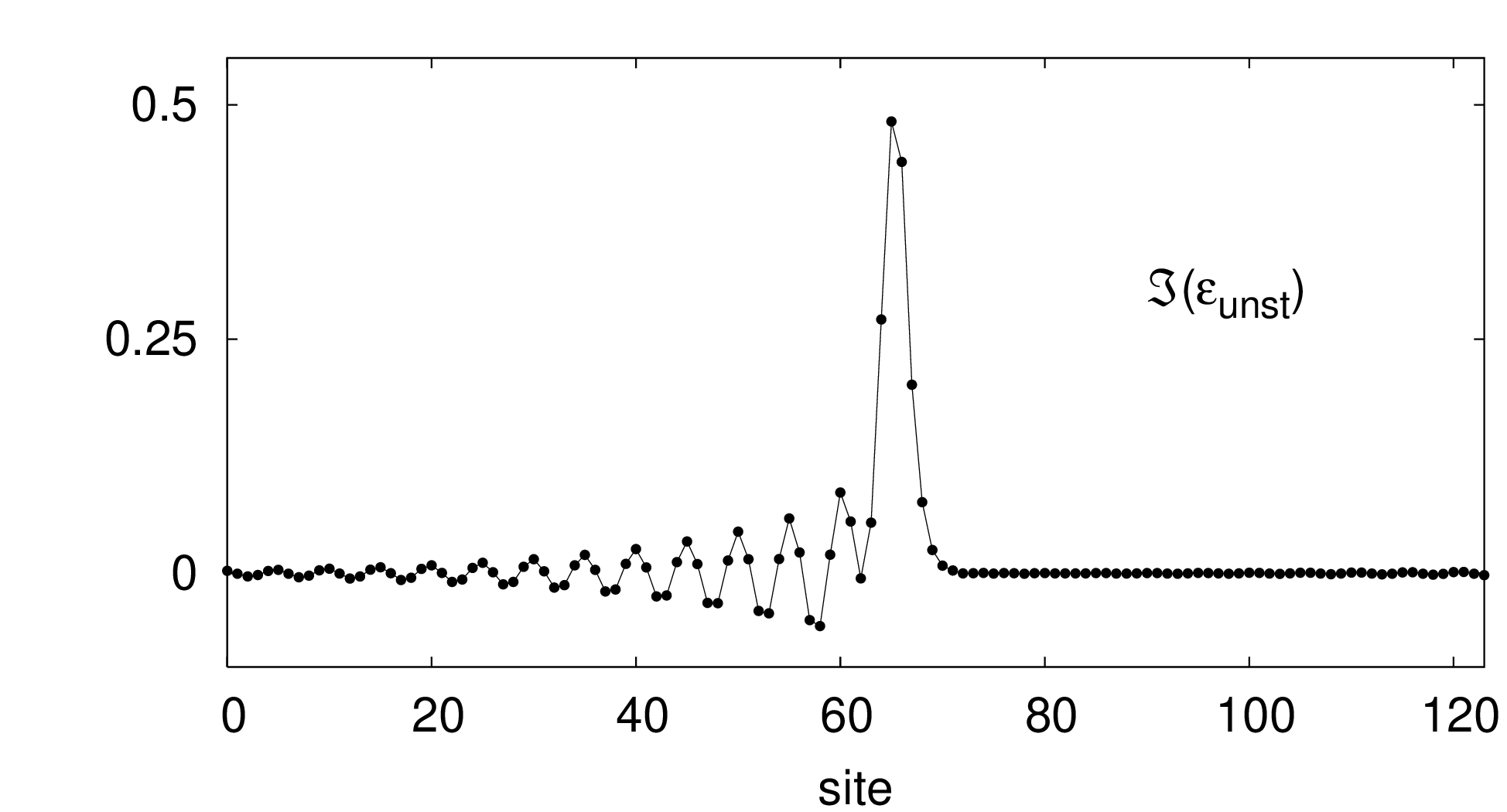}
}}
\end{tabular}
\caption{{\bf (a)} Floquet angle evolution of the spectra of a $(1,1)$
resonant breather with $\omega_{b}=2.678$. The thick trajectory
corresponds to the localized eigenvector that becomes unstable 
($\theta_{Floq}=0$ interval). Instantaneous profile of the real 
{\bf (b)} and imaginary {\bf (c)} part of the Floquet unstable 
eigenvector of a $(1,1)$ breather with $\omega_{b}=3.207$ and
$\nu=0.26$. The decaying tails along the direction opposite to the 
motion reveals the energy loss that the unstable eigenvector causes 
to the solution.} 
\label{fig:uns_prof}
\end{figure}

We turn now to {\em background instabilities}. Once we know the plane
wave content ($k_{0}$, $k_{1}$,..) of a ($p/q$)-resonant fixed
point, we can know whether the solution is subject to MI or not
and, if it is unstable, what are the harmful perturbations ($Q$).
This problem is not so simple because we cannot know {\em a
priori} the plane wave content if we do not have the amplitudes of
each one (\ref{resonant}). However, we can derive a necessary
condition for not having MI if we consider that, from
(\ref{resonant}), the background is always composed of at least 
one plane wave ($m=0$) with $k_{0}$ between $\lbrack -\pi
/2,0\rbrack$. From this we can simplify the analysis of the
background stability to the $k_{0}$ plane wave stability as a
necessary condition for the MB stability. For this we calculate,
for each $\nu$ and $k$,  the value of the right-hand side of
(\ref{disp_pert}) for all the range of $Q$ ( $\lbrack -\pi, \pi
\rbrack$ ) and $A$. If this value is always positive the plane
wave with this $k_0$ is free from modulational instabilities at
this point of the model (\ref{Salerno}) with parameter $\nu$. From
this extensive exploration we obtain, see Fig. \ref{fig:noMI}, the region in
the $k-\nu$ plane where MI is present.
\begin{figure}[!tbh]
\begin{tabular}{cc}
\centerline{\resizebox{8.cm}{!}{%
\includegraphics[angle=-90]{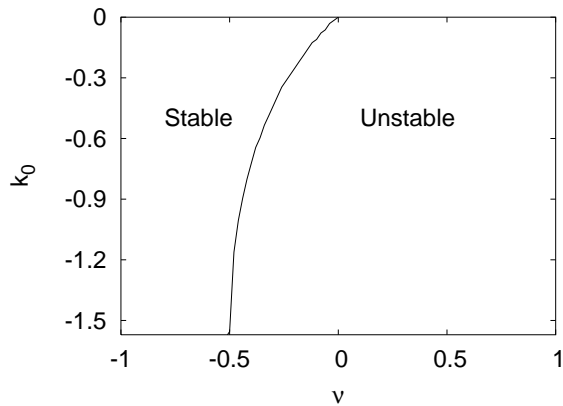}
}}
\end{tabular}
\caption{Modulational instability existence diagram for a plane wave
with wave number $k_{0} \in [-\pi/2,0]$. This diagram fixes the
region where mobile discrete breathers with a background composed of only one
plane wave do not suffer from background instability.} 
\label{fig:noMI}
\end{figure}

In the range of $\nu$ between $\lbrack -1,-0.5 \rbrack$ there is
no modulational instability for single plane waves of any value of
$k$ between $\lbrack -\pi/2,0 \rbrack$, and in particular for
$k_{0}$. However, this does not guarantee that moving breathers
are free from these instabilities in this region, unless the
background has only one plane wave (as is sometimes the case). On
the contrary, in the region $\nu>0$ any moving breather suffers
such instabilities. The transition area in the region $\nu \in
\lbrack -0.5,0 \rbrack$ presents MI depending on which $k_{0}$ we
have. For the range where no plane-wave with $k$ between $\lbrack
-\pi/2,0 \rbrack$ is subject to MI we can assure that if there
is only one contribution, $k_{0}$, to the background the
corresponding MB solution is stable. For example, this is the case
for $(1/1)$ resonant breathers if  $\omega_{b}>4$ and for $(1/2)$
resonant breathers if $\omega_{b}>8.46$. The Floquet spectra of a 
moving breather satisfying these requirements is plotted in 
Fig. \ref{fig:Floquet}.c.

\begin{figure}[!tbh]
\begin{tabular}{cc}
\centerline{ {\bf (a)}
\resizebox{5.cm}{!}{%
\includegraphics[angle=-90]{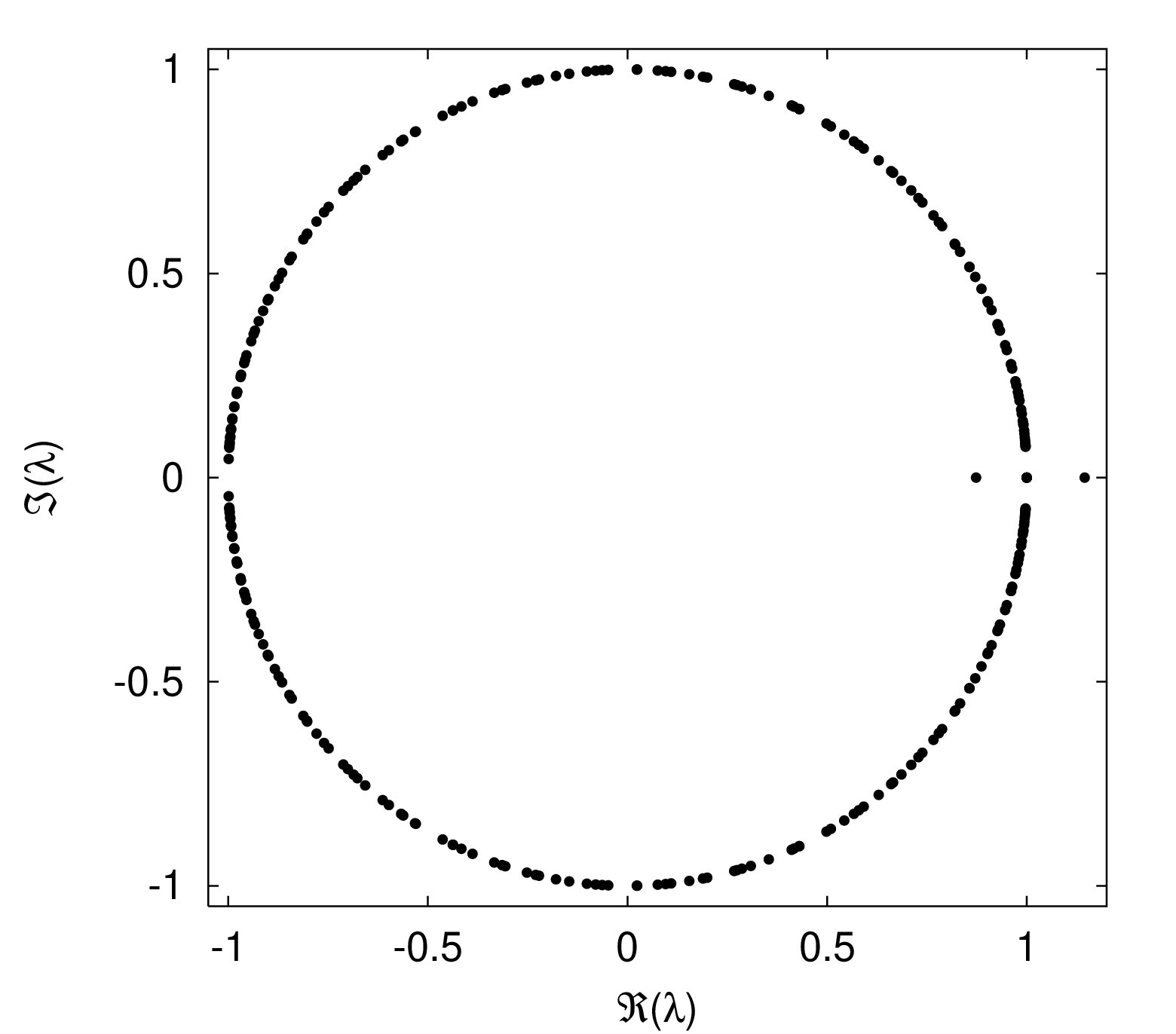}
}
{\bf (b)}
\resizebox{5.cm}{!}{%
\includegraphics[angle=-90]{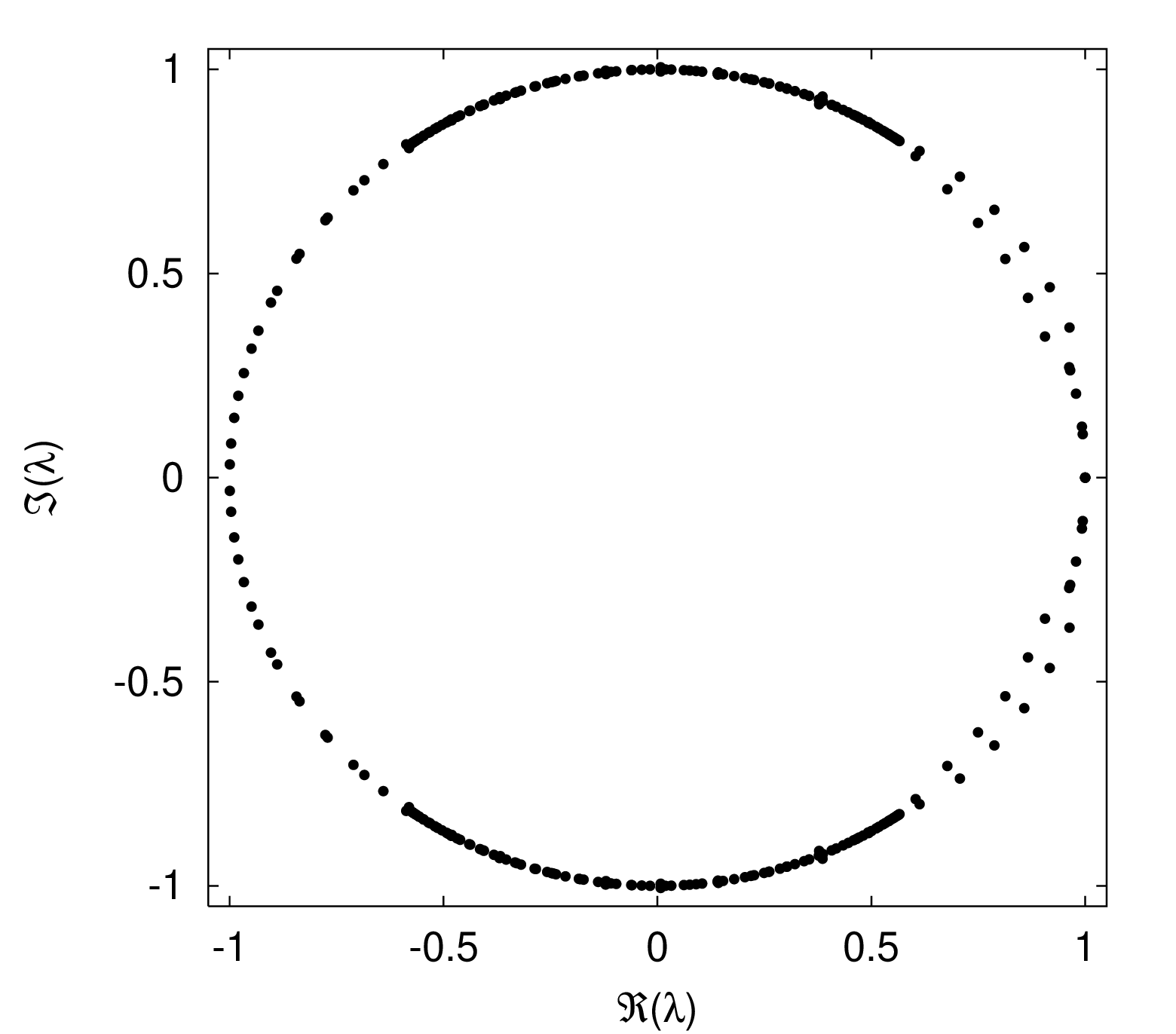}
}
{\bf (c)}
\resizebox{5.cm}{!}{%
\includegraphics[angle=-90]{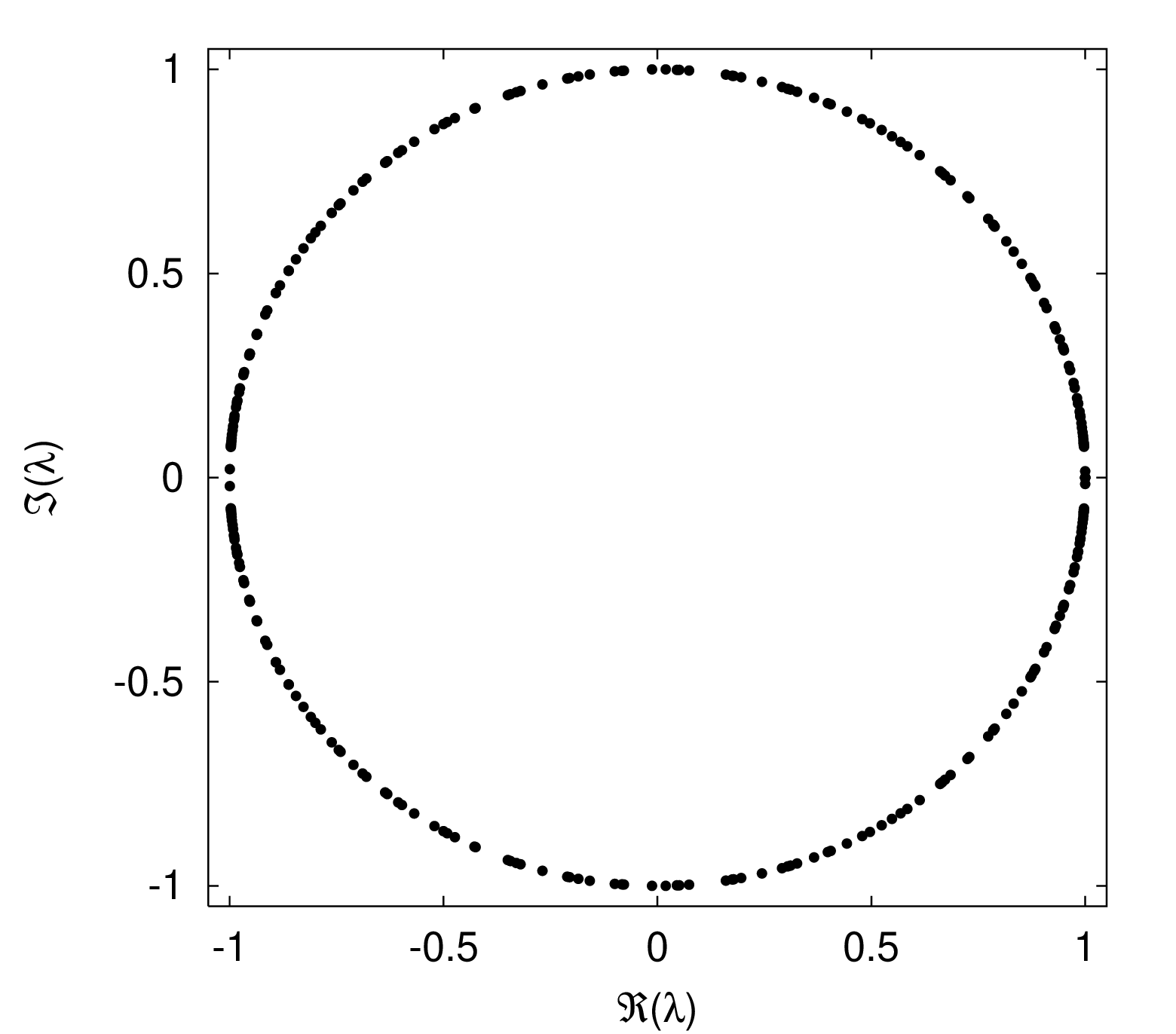}
}}
\end{tabular}
\caption{
Floquet spectra of $(1,1)$ resonant breathers: {\bf (a)} for
$\omega_{b}=4.348$, $v_{b}=0.692$ and $\nu=0.08$ the spectra shows the
core (localized) instability; {\bf (b)} for $\omega_{b}=6.610$,
$v_{b}=1.052$ and $\nu=0.07$ the spectra shows the background
(modulational) instability (also present but not visible 
in {\bf (a)}); {\bf (c)} for $\omega_{b}=4.348$, $v_{b}=0.692$ 
and $\nu=-0.39$ the solution is linearly stable.}
\label{fig:Floquet}
\end{figure}

After the analysis of both types of instabilities eventually
experienced by moving Schr\"odinger breathers, we finally report on a
most relevant numerical fact revealed by the Floquet analysis of
the family of {\em immobile discrete breathers} for $\nu < 0$: Near $\nu
\simeq -0.3$ an immobile two-site DB experiences a mirror
symmetry-breaking (pitchfork) bifurcation becoming linearly
unstable. When approaching the bifurcation point, two conjugate
Floquet eigenvalues quickly approach $+1$, where they meet, and
then separate along the real axis. The eigenvector associated to
the unstable $\lambda > 1$ Floquet eigenvalue is localized and
odd-symmetric, and is termed the symmetry-breaking or depinning
mode $\phi^{dep}$. We recall here that the background of an
immobile breather is the rest state $\hat{\Phi} = 0$, whose continuous
spectrum consists of small amplitude (linear) plane waves. The
depinning mode, on the other hand, is a localized core instability
of the immobile breather, favoring a translation of the core
center. For a smaller value of $\nu\simeq -0.39$ there is another
symmetry-breaking bifurcation where the two-site breather becomes
stable, again interchanging the stable character with the
one-site. The corresponding bifurcation diagram for these two symmetry
breaking transitions is plotted in Fig. \ref{fig:SSB}.  

\begin{figure*}[!tbh]
\begin{tabular}{cc}
\centerline{
\resizebox{16.cm}{!}{%
\includegraphics[angle=-0]{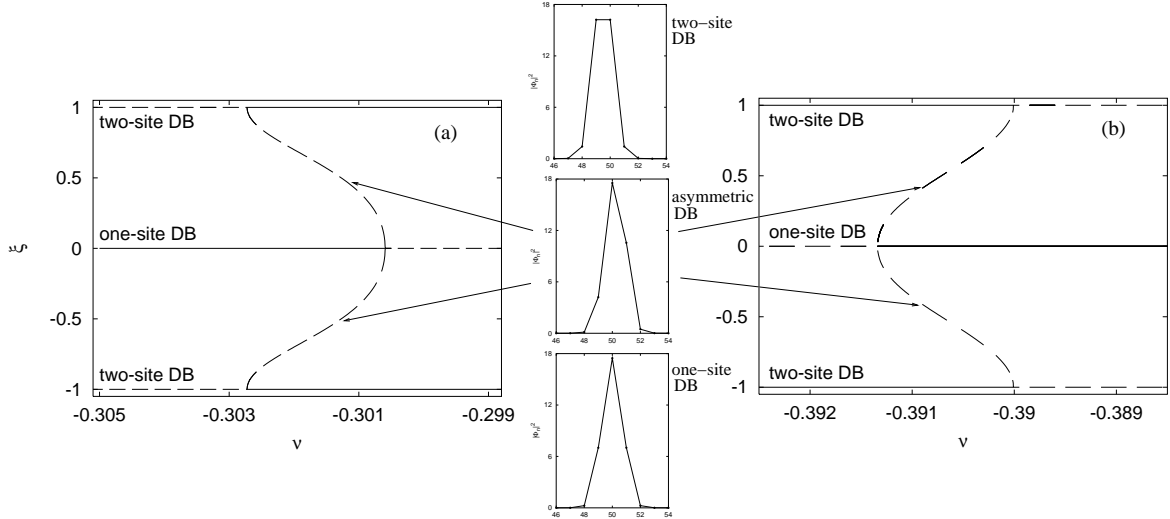}
}
}
\end{tabular}
\caption{Graphical representation of the two first {\em symmetry breaking
bifurcactions} for $\nu<0$. The quantity $\xi$ in the vertical axes
of both figures is defined, referred to the one-site breather, as the
difference between the modulus $|\Phi|$ of the two sites adjacent 
to the maximum ($|\Phi_{max}|$), {\em i.e.} $\xi \sim
|\Phi_{max-1}|-|\Phi_{max+1}|$. For the one-site DB $\xi=0$ and for
the two-site DB $\xi=1$, for this $\xi$ is conveniently normalized
with the the difference between $\Phi_{max}$ and $\Phi_{max\pm
1}$. The continuous lines represent the regions where the static
solutions are linearly stable while the discontinuous ones represent the
unstable regions. The modulus profile of the three immobile coexisting
solutions are plotted in the central insets for $\omega_{b}=6.215$ 
and $\nu=-0.3012$.} 
\label{fig:SSB}
\end{figure*}

In the first symmetry breaking bifurcation, two unstable 
mirror-asymmetric immobile breathers emerge from the bifurcation point, 
progressively evolve toward the (stable)
two-site breather, and finally collide in a new pitchfork
bifurcation from where a unstable two-site breather emerges. The net
result is an inversion of stability between one-site and two-site
immobile breathers. Around the narrow interval of $\nu$ values
where these two bifurcations occur, the energies of the three
types of breathers involved (one-site, two-site, and asymmetric)
have very small differences. From a particle perspective, this
should make the breather motion easier. It is precisely in this
same narrow interval where (see \ref{Background.Amplitude}) we
observe that the background amplitude of moving breathers becomes
negligible. This is not a coincidence as we will argue in 
\ref{sec:Particle}.

\section{Particle perspective on discrete breathers}
\label{sec:Particle}

The appealing framework and success of collective variable
approaches (see e.g. \cite{Vakhnenko,Claude,Cai,Kundu,MackaySep}) 
to the problem of nonintegrable
motion of discrete breathers relyes on the fidelity of a
particle-like description of these field excitations that they
provide. In these approaches, the effective dynamics of only a few
degrees of freedom (e.g. the localization center, and the spatial
width of the state, etc...in some instances
\cite{Smerzi}) replaces the whole description of the
moving localized state.

Though unable to account for all the nonintegrable
features, perturbative collective variable theories of NLS
lattices provide a sensible physical characterization of important
features of the nonintegrable mobility of localized solutions, like
the emergence \cite{KivsharCampbell} of a Peierls-Nabarro barrier to motion.
Here we summarize the main results of this particle-like
description and compare them with the behaviour of numerically
exact ($p$, $q$) resonant moving breathers. Our goal is twofold:
to acquire a correct physical understanding of the
numerical facts, and then to make an assesment of validity and
intrinsic limitations of collective variable approaches.

\subsection{Collective variables theory.}

A presentation of the particle perspective on moving Schr\"odinger
breathers near the A-L integrable limit can be found in
\cite{Cai} (see also \cite{Claude,Vakhnenko,Kundu,MackaySep}),
where the interested reader will find the relevant formal aspects
of the theory.

Using the integrable solitary wave (\ref{A-Lbreather}) as an
ansatz for the moving breather solution in the perturbed A-L
lattice, $\nu \not=0$ and small in (\ref{Salerno}), one considers
the parameters $\alpha$, $\beta$, $x_{0}$ and $\Omega$ as
dynamical variables (variation of constants). The time evolution
of these parameters in the perturbed lattice is governed by:
\begin{eqnarray}
\dot{x}_{0}&=&2\sin\alpha\frac{\sinh\beta}{\beta}
\label{Coll_Var1}
\\
\dot{\Omega}&=&2\cos\alpha\cosh\beta+\alpha\dot{x}_{0}+g(\beta)
\label{Coll_Var2}
\\
\dot{\beta}&=&0 \label{Coll_Var3}
\\
\dot{\alpha}&=&
-\nu\sum_{s=0}^{\infty}\frac{8\pi^{3}\sinh^{2}\beta}
                            {\beta^{3}\sinh({\pi^2 s/\beta})}
\sin(2\pi sx_{0}) \label{Coll_Var4}
\end{eqnarray}
where
\begin{eqnarray}
g(\beta)&=&2\nu\lbrack
\frac{2\sinh\beta\cosh\beta}{\beta}-\frac{\sinh^{2}\beta}
{\beta}-1\rbrack
\nonumber
\\
&+&\nu\sum_{s=1}^{\infty} 4\pi^{2}\cos(2\pi sx_{0}) \lbrack
\frac{\sinh^{2}\beta\cosh({\pi^2
s/\beta})\pi^2s}{\beta^{4}\sinh^2({\pi^2 s/\beta})} \nonumber
\\
&-&\frac{2\sinh^{2}\beta}{\beta^{3}\sinh({\pi^2 s/\beta})}
+\frac{2\sinh\beta\cosh\beta}{\beta^{2}\sinh({\pi^2 s/\beta})}
\rbrack \label{g(b)}
\end{eqnarray}

These relations can be viewed as the Euler-Lagrange equations of
the collective variable Lagrangian obtained in \cite{Cai}. The
variation of the breather parameters give the evolution of
solution (\ref{A-Lbreather}) for the perturbed A-L equation.
Furthermore, one can regard eqs. (\ref{Coll_Var1}) and
(\ref{Coll_Var4}), as the Hamilton equations for the canonical
conjugate variables $x_{0}$ and $\alpha$ of the following
effective hamiltonian:
\begin{eqnarray}
{\cal H}_{\rm eff}&=&{\cal T}_{\rm eff}+{\cal V}_{\rm
eff}=-2\cos\alpha\frac{\sinh\beta}{\beta} \nonumber
\\
&-&\nu\sum_{s=1}^{\infty}\frac{4\pi^{2}\sinh^{2}\beta}
                     {\beta^{3}\sinh({\pi^2 s/\beta})}
                             \cos(2\pi sx_{0})\;,
\label{CV_Ham}
\end{eqnarray}

This effective hamiltonian dictates the dynamics of the position
of the solitary wave. Note that the (collective) variable $\beta$
is an invariant of motion, so it enters as a parameter into the
effective hamiltonian, and that the time-average value of
$\dot{\Omega}$ (the parameter $\omega_b$ of the integrable
solitary wave, now a function of time) is an increasing function
of this parameter $\beta$. The effective potential ${\cal V}_{\rm
eff}$ acts as a barrier to the displacement motion ($x_{0}$
variations) and is naturally related to the Peierls-Nabarro
potential. The amplitude of this barrier is an increasing function
of both the nonintegrability parameter $|\nu|$ and $\beta$. The
equilibrium points (representing immobile breathers) of this
potential are $x_{0}=n$ and $n\pm1/2$ with $n$ an integer. For
$\alpha=0$, the former are stable (centers) one-site breathers,
while the latter are unstable (saddle) two-site breathers; for the
case $\alpha=\pi$ (staggered breathers) the stability is reversed.

A remarkable further consequence is the following \cite{Claude}:
{\em  there are no perturbative traveling wave solutions, for
values of $\nu$ larger than certain critical value
$\nu_{cr}(\beta)$}. In particular, for $\beta > \beta_c \simeq
3.6862$ , one cannot continue A-L mobile breathers ({\em i.e.}
$\nu_{cr}=0$ (see also important remarks in \cite{MackaySep}).
This consequence could be also (qualitatively) expected for a
class of nonintegrable Schr\"odinger lattices (for some qualified
perturbations of the integrable limit) with on-site nonlinearity.
One expects also that lattices with purely inter-site (FPU-like)
nonlinearity do not show this kind of transition. 

In Fig. \ref{fig:Claude} we plot typical phase portraits at both
sides of $\nu_{cr}$. Fig. \ref{fig:Claude}.a shows the dynamics
for $\nu$ smaller than the treshold value (given by $\beta$): 
there are open trajectories in $x_{0}$ corresponding to mobile 
breathers and closed orbits between the separatrix manifolds 
corresponding to breathers which oscillate around the equilibrium 
position of ${\cal V}_{\rm eff}$. Fig. \ref{fig:Claude}.b is the 
phase portrait after the transition: there are no longer mobile
solutions and (besides the oscillating breathers) there are
instead open trajectories in $\alpha$. The transition point,
for a given $\beta$, occurs when trajectories with rotating
$\alpha$ appear, and moving breathers disappear as the effect of
separatrix line rearrangement on the cylinder ($x_0$, $\alpha
({\mbox {modulo}} 2\pi)$) phase portrait.

\begin{figure}[!tbh]
\begin{tabular}{cc}
\centerline{ {\bf (a)}
\resizebox{8.cm}{!}{%
\includegraphics[angle=-90]{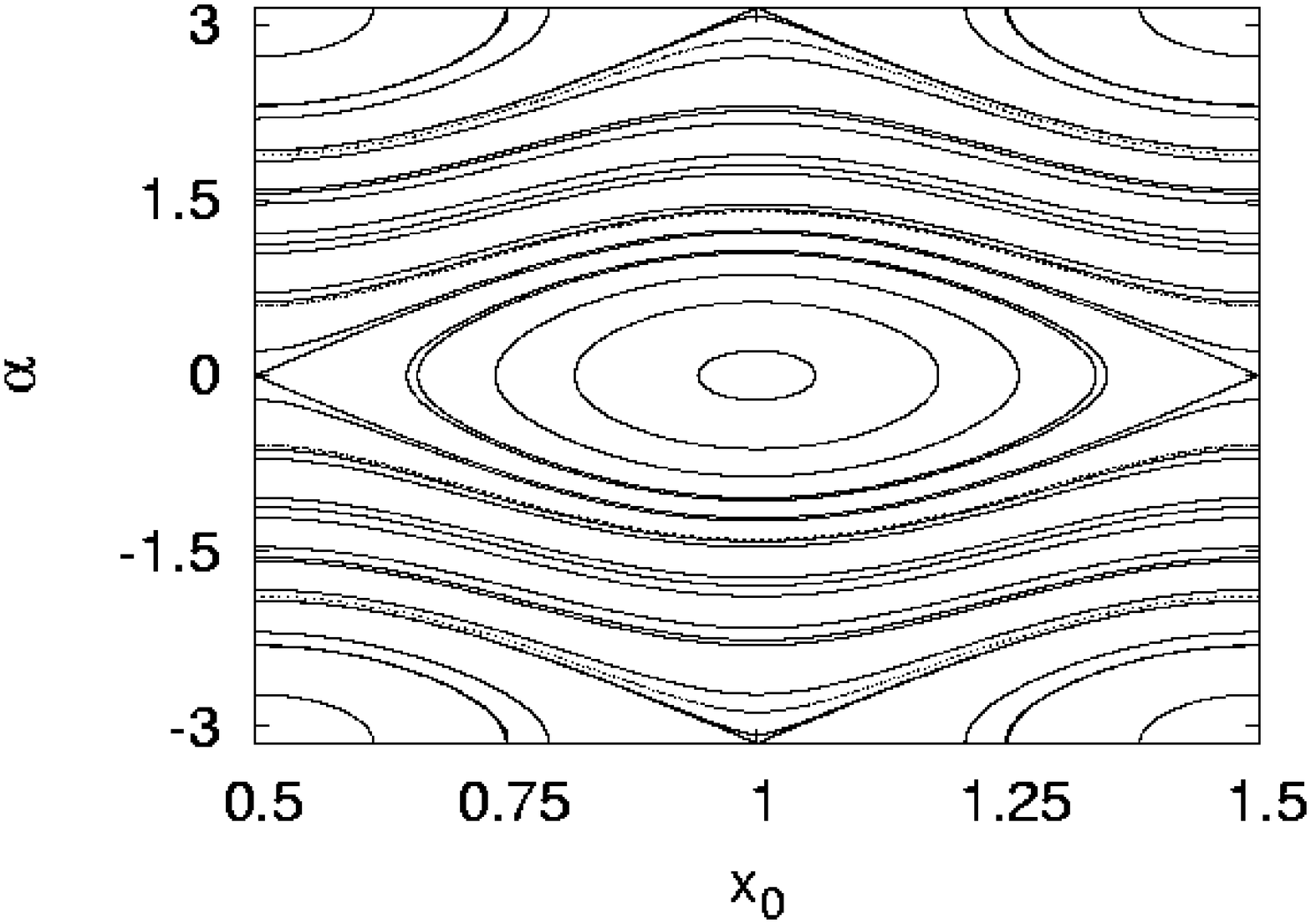}
}
{\bf (b)}
\resizebox{8.cm}{!}{%
\includegraphics[angle=-90]{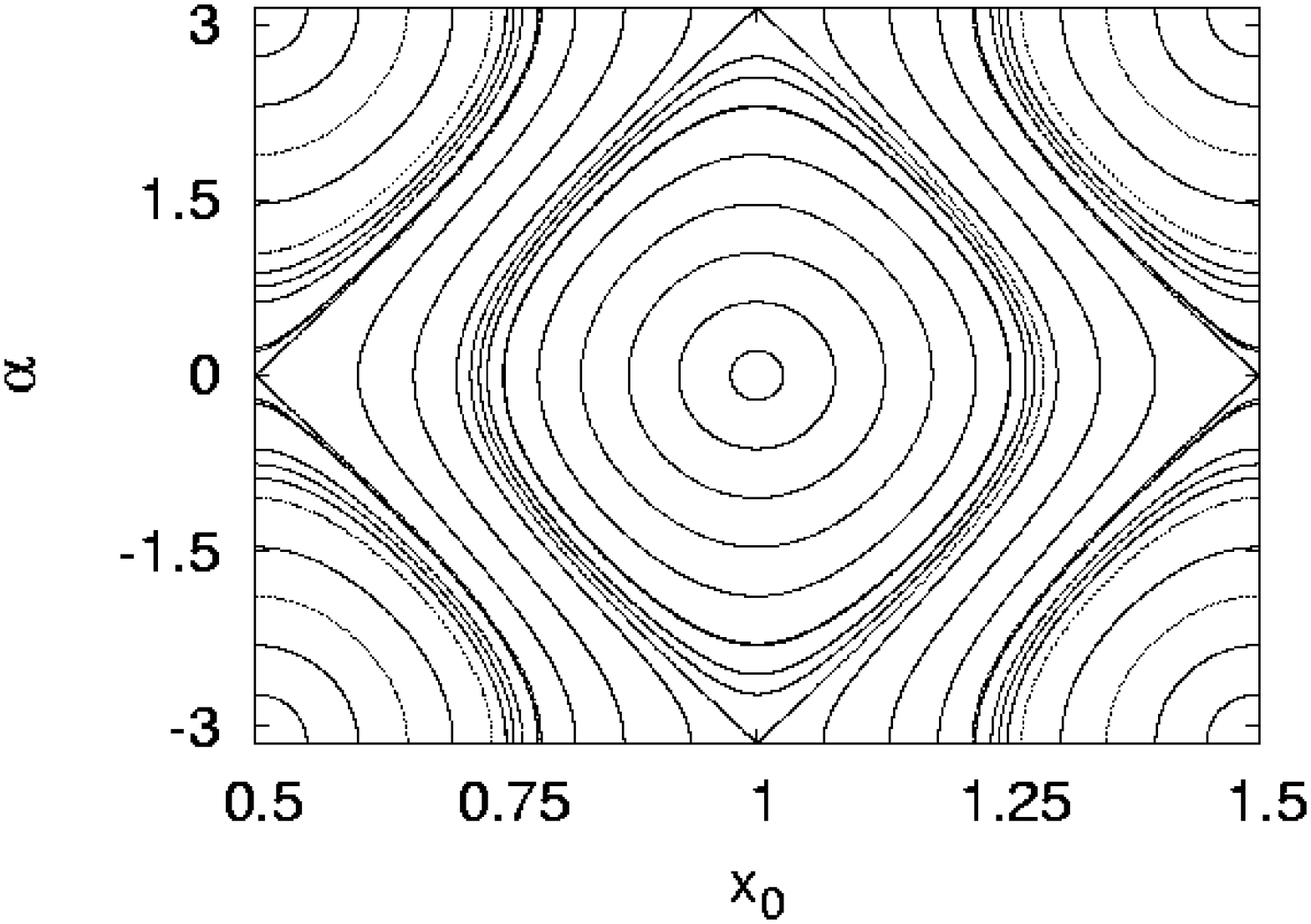}
}}
\end{tabular}
\caption{Collective variable $(\alpha, x_{0})$ phase portrait
transition for a value of $\beta=3.0$. {\bf (a)} Shows the phase
portrait for $\nu=0.2\quad(<\nu_{cr})$, there are $x_{0}$-unbounded
trajectories (mobile breathers) coexisting with bounded ones
(oscillating breathers). When $\nu=1.0\quad(>\nu_{cr})$ {\bf (b)} we 
only have $x_{0}$-bounded trajectories: there are no mobile solutions.}
\label{fig:Claude}
\end{figure}

Note that the existence of oscillating breathers is a consequence
of the existence of a Peierls-Nabarro potential. These breather
solutions do not perturbatively continue from the integrable
limit. In \ref{Oscillating} we will investigate them and provide
further numerical confirmation of the existence of these genuinely
nonperturbative solutions, predicted by the collective variables
theory.

\subsection{Energy balance governs mobility.}
\label{PN-BARRIER}

In order to correlate collective variable predictions with the
numerical results presented in section \ref{sec:Mobile} one should
first realize that our direct numerical approach computes
breathers with fixed values of $\omega_b$ and $v_b$ and that these
parameters are not tied to any specific ansatz. In particular, the
connection of these two parameters with the collective variables
is given by eq. (\ref{omegav}) in the integrable limit. For
the perturbed (near-integrable) lattice, $\omega_b$ and $v_b$ are
identified as the time averages of $\dot{\Omega}$ and
$\dot{x}_{0}$, respectively.

The Peierls-Nabarro (PN for short) barrier is naturally
identified as the energy difference (given by the Hamiltonian
(\ref{Salerno_Ham})) between the two immobile breathers of the
same frequency $\omega_{b}$, one centered at a site $n$ and the
other (two-site) at a bond $n\pm1/2$ :
\begin{equation}
E_{PN}(\nu,\omega_{b})=H(\nu,\omega_{b},n)-H(\nu,\omega_{b},n\pm\frac{1}{2})
\label{PN}
\end{equation}
In the integrable A-L limit this barrier is zero due to
the degeneracy (continuous translation invariance) of the breather
family solution, but for $\nu\not=0$ this invariance is broken and
only these two isolated solutions persist. The energy difference
of the two pinned solutions is thus viewed as the minimal extra
``{\em kinetic energy }'' of center of mass translation that a
mobile breather must incorporate for overcoming the barriers to
its motion.

\begin{figure}[!tbh]
\begin{tabular}{cc}
\centerline{\resizebox{8cm}{!}{%
\includegraphics[angle=-0]{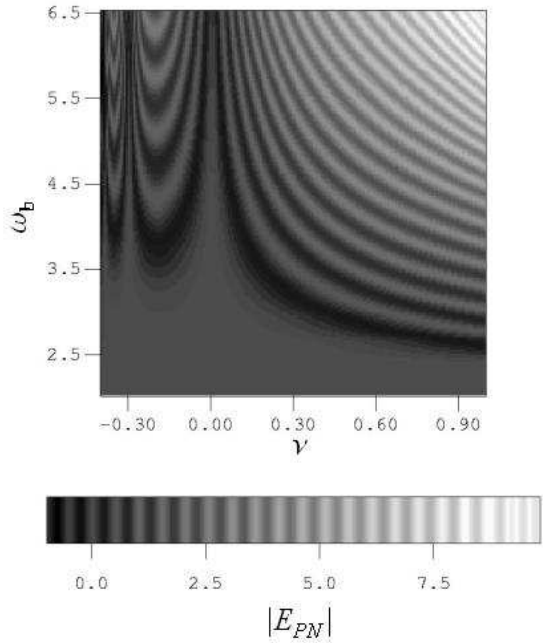}
}}
\end{tabular}
\caption{Density Plot of the absolute value of the Peierls Nabarro
barrier, $|E_{PN}|$, as a function of $\omega_{b}$ and $\nu$. For
positive values of $\nu$, $|E_{PN}|$ is a monotonous increasing function of
$\nu$ and $\omega$. For negative values the plot reveals the
oscillating behaviour of $|E_{PN}|$ as a function of $\nu$ (for a given 
value of $\omega_{b}$).  
} 
\label{fig:PN}
\end{figure}

We have studied the behavior of the PN barrier in the Salerno
model by continuing immobile breathers, both centered at a site
and at a bond, while computing their energy difference. The
computations of the barrier are made for a grid of values of
$\omega_{b}$. Fig. \ref{fig:PN} shows the ``{\em equipotential}''
lines of the PN barrier in the ($\nu$, $\omega_{b}$) plane. The
results show different behaviors depending on the sign of $\nu$:

$\nu < 0$. Here one observes the effects of the symmetry-breaking
bifurcations cascade described in \ref{Flo.An}. The successive
stability inversions between site and bond centered breathers
involve a substantial decrease of the Peierls barrier. The
appearance of asymmetric solutions in these bifurcations
introduces a new energy and, correspondingly, the Peierls barrier
is computed as the maximum energy difference between the three
pinned solutions: the two symmetric (site and bond centered) and
the asymmetric breather. 

$\nu > 0$. Here the behavior of the Peierls barrier
follows qualitatively the collective variable predictions on the
effective potential experienced by the particle. The increasing
character, with $\nu$ and $\omega_{b}$, of the numerical barrier
is qualitatively the same as that predicted from $V_{eff}$ (as a
function of $\nu$ and $\beta$) by the theory.

The PN barrier of ($\omega_{b}$) immobile breathers and the 
background amplitude of ($\omega_{b}$, $v_{b}=\frac{p\omega_{b}}
{2\pi q}$) mobile breathers are in fact strongly
correlated. This correlation is obtained considering the functions  
$|E_{PN}|(\nu)$ and $|\hat{\Phi}_{backg}|^2(\nu)$. Both functions 
are plotted for a fixed value of $\omega_{b} = 4.34$ in
Fig. \ref{fig:AyPN}.a. The behavior of $|E_{PN}|(\nu)$ for negative
$\nu$ (revealing the cascade of bifurcations explained before in
\ref{Flo.An}) is closely followed by $|\hat{\Phi}_{backg}|^2(\nu)$ with
the corresponding sequence of growths and decays. The strong
correlation holds also for positive values of $\nu$, where
numerical PN barrier data are available for a larger interval of
$\nu$ values (due to the absence of the symmetry-breaking cascade
of bifurcations). Indeed, the correlation is so strong that one is
tempted to view the PN barrier and the background amplitude as
complementary aspects of a single phenomenon: the breaking of the
continuous translational invariance, and the associated lack of
core momentum conservation \cite{note}. Indeed, the background amplitude of
moving breathers is a monotone increasing function of the PN
barrier of pinned breathers of the same frequency, as shown in
Fig. \ref{fig:AyPN}.b, where $|\hat{\Phi}_{backg}|^2(|E_{PN}|)$ is plotted.

However, we also observe clearly in Fig. \ref{fig:AyPN}.a that,
when the continuation end is approached, the rate of growth of
$|\hat{\Phi}_{backg}|^2(\nu)$ increases dramatically (the concavity of
the curve in log scale turns upwards), while the PN barrier does
not increase much faster than before. This is reflected in figure
\ref{fig:AyPN}.b, where the slope approaches verticality, indicating
that, in this range of $E_{PN}$ values, the background grows rapidly.

\begin{figure*}[!tbh]
\begin{tabular}{cc}
{\bf (a)}
\resizebox{8cm}{!}{%
\includegraphics[angle=-90]{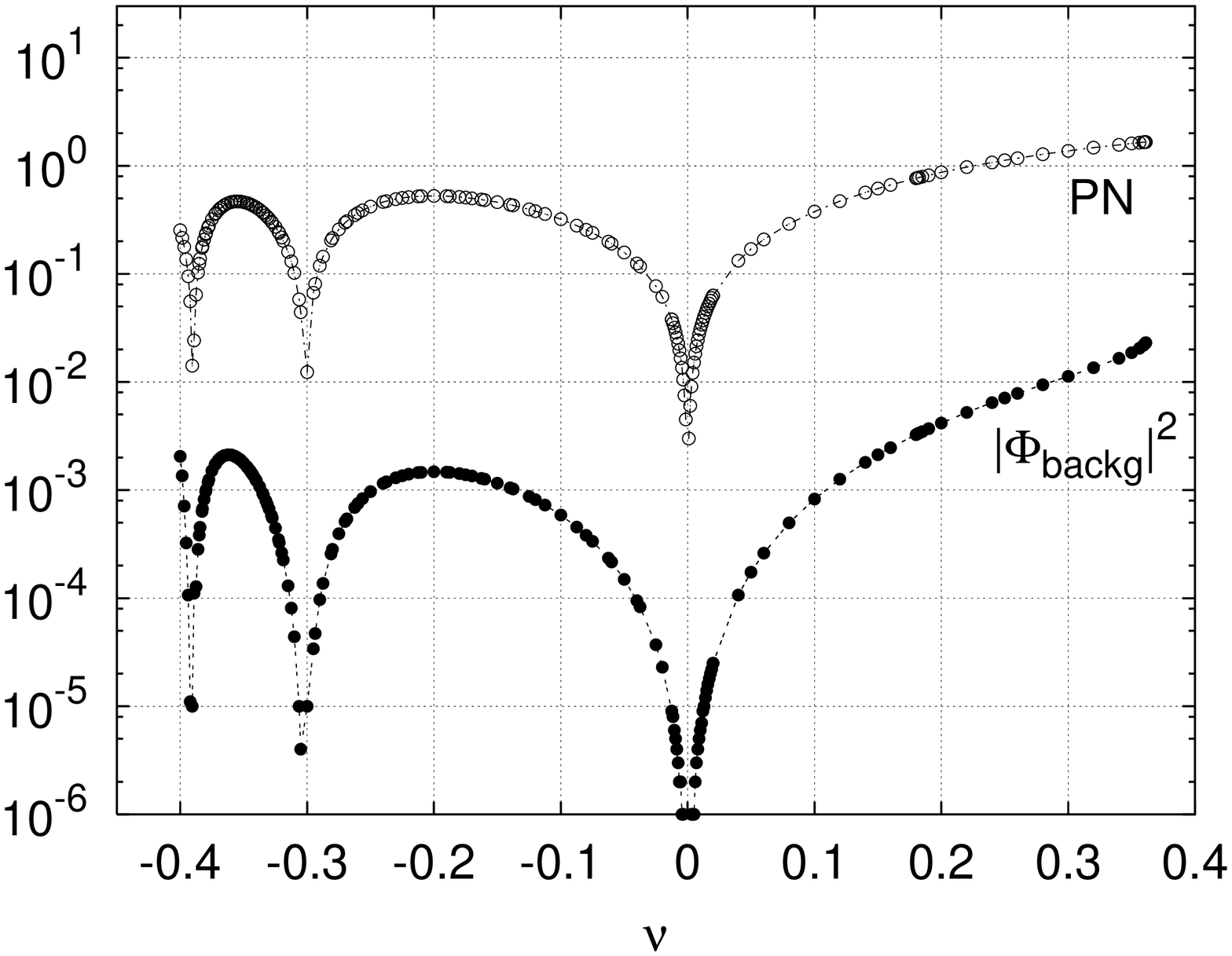}
}
& {\bf (b)}
\resizebox{8cm}{!}{%
\includegraphics[angle=-90]{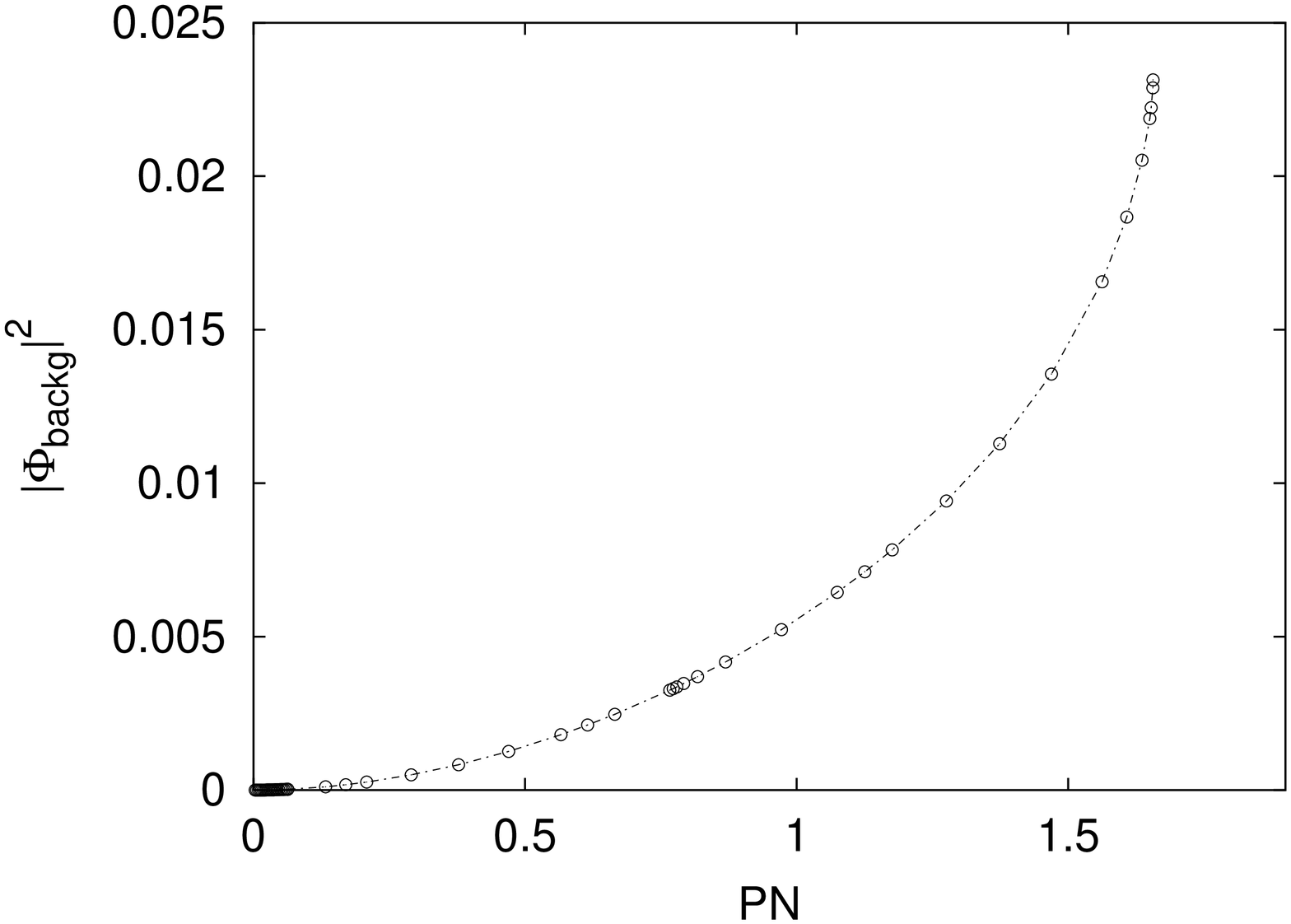}
}
\end{tabular}
\caption{Peierls Nabarro barier $|E_{PN}|$ from inmobile breather with
$\omega_{b}=4.34$ and background square amplitude $|\Phi_{backg}|^{2}$ for
a $(1,1)$ resonant breather of the same frequency ($v_{b}=0.691$). In 
{\bf (a)} we show both quantities in semi-log scale as functions of $\nu$,
illustrating the strong correlation between them for both signs of
$\nu$. Figure {\bf (b)} shows for positive values of $\nu$, the
nonlinear relation between $|\hat{\Phi}_{backg}|^{2}$ and $|E_{PN}|$. Note
the sudden increase of the slope close to the end of numerical continuation.} 
\label{fig:AyPN}
\end{figure*}

This numerical observation suggests taking a closer look at the precise
influence of the background amplitude on the core energy variations
associated with the existence of PN barriers. To this end, we
use the conservation of the Hamiltonian (\ref{Salerno_Ham}) and insert this
equation into the form (\ref{core-bckg}) of the ($p$, $q$) resonant fixed
point. The energy of the solution can be decomposed in the following terms:
\begin{equation}
{\cal H}[\hat{\Phi}]={\cal H}[\hat{\Phi}_{core}]+ {\cal H}[\hat{\Phi}_{backg}]+{\cal
H}^{int}\;,
\end{equation}
where ${\cal H}^{int}$ is the interaction energy, {\em i.e.} the
crossed terms of $\hat{\Phi}_{core}$ and $\hat{\Phi}_{backg}$. Let us now
consider the simplest case in which the background has a single
resonant plane wave. Along with the total energy, also the energy
of the plane wave is a constant in time so that
\begin{equation}
\frac{\partial{\cal H}[\hat{\Phi}_{core}]}{\partial
t}=-\frac{\partial{{\cal H}^{int}}}{\partial t}\;.
\label{Energy_balance}
\end{equation}
In other words, the variations of the core energy during the
motion are exactly compensated by those of the interaction term.

If one takes, as an ansatz for $\hat{\Phi}_{core}$, the A-L solution, one
formally obtains for ${\cal H}^{core}\equiv{\cal H}[\hat{\Phi}_{core}]$ 
the collective variables Hamiltonian (\ref{CV_Ham}). But note that
here it would not anymore be a constant of motion, due to the
interaction with the background. Instead (as P. Kevrekidis suggested
to us), we directly compute numerically the evolution of the core
energy ${\cal H}^{core}$, which in turn determines ${\cal H}^{int}$ up to
an additive constant.

For this we take a fixed point solution with a single plane wave
in its background, and then substract off the plane wave to obtain
$\hat{\Phi}_{core}$, from where ${\cal H}^{core}(t)$ is computed. In Fig.
\ref{fig:Hint} we have plotted the evolution of the core energy as
a function of the localization position (center) of the breather
core. The localization center of a lattice function $\Phi_n$ is
defined using the conserved norm (\ref{norm}):
\begin{equation}
x_{0}=\frac{\sum_{n}{n\ln(1+\mu |\Phi_{n}|^2)}}{\mu{\cal N}}\quad.
\label{loc_cent}
\end{equation}
As expected, the core has extracted the maximum available from the
interaction energy (with the background) when the core passes at
$n\pm1/2$ (maxima of the PN barrier) and has returned it to the
interaction term when at $n$ (minima of the PN barrier).

\begin{figure}[!tbh]
\begin{tabular}{cc}
\centerline{
\resizebox{8cm}{!}{%
\includegraphics[angle=-90]{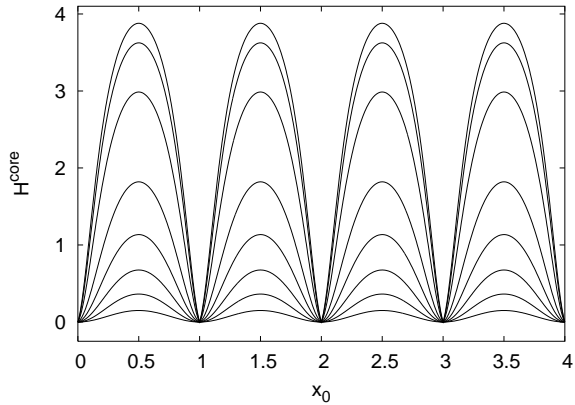}
}}
\end{tabular}
\caption{Plot of ${\cal H}^{core}$ of a $(1,1)$ resonant breather 
as a function of the localization center $x_{0}$ for different 
values of $\nu$. The parameter of the solution are $\omega_{b}=5.056$ 
and $v_{b}=0.805$. The values of $\nu$ are $0.04$, $0.08$, $0.12$, $0.16$,
$0.20$, $0.24$, $0.25$ and $0.2512$ (end of the continuation), the
amplitude of the oscillation of ${\cal H}^{int}$ grows with $\nu$. (The
minimum value of ${\cal H}^{int}$ has been set to zero in order 
to compare the differents functions.)} 
\label{fig:Hint}
\end{figure}

Another interesting feature of these numerically obtained functions 
is seen from the variations in the form of the oscillation of 
${\cal H}^{core}$ as the nonintegrable parameter $\nu$ is increased. 
At the same time, as the energy difference between $n$ and 
$n\pm1/2$ increases the modulus of the derivative $\partial{\cal H}
^{core}/\partial x_{0}$ in the neighborhood of  $x_{0}=n$ also
increases. These variations become faster when the end of the
continuation is approached, reaching a cuspidal point for the last 
$\nu$ reached. The background amplitude is included in ${\cal
H}^{int}$, and of course in $\partial{\cal H}^{int}/\partial t$; 
the dramatic variation of it at the end of the continuation could 
be interpreted in terms of this derivative variation in $x_{0}=n$.

\subsection{Oscillating breathers}
\label{Oscillating}

The emergence of the Peierls barrier and the behavior of the
background amplitude illustrate the physical interpretation of this
background as a (p/q)-resonant energy support to overcome the
barrier to motion. We now confirm this statement searching for
another kind of solution: {\em oscillating breathers}. These
solutions are predicted by collective coordinates approaches and
are a consequence of the loss of translational invariance out of
the integrable limit. Following the above interpretation of the
background role one can imagine certain solutions with a
background amplitude not high enough for surpassing the Peierls
barrier and allowing travel along the lattice. In terms of a well
defined potential, considering the particle perspective, the
center of these localised solutions would be oscillating between
$(n-1/2)$ and $(n+1/2)$ for the unstaggered ones or between $n$
and $(n\pm1)$ for the staggered ones.

From our perspective, the oscillating breathers are solutions with
two frequencies: the internal one of the breather ($\omega_b$) and
the one corresponding to the oscillatory motion ($\omega_{osc}$).
Once again, we have a problem dealing with two time scales and
consequently we have to impose that the two frequencies are
commensurate $p\omega_{b}=q\omega_ {osc}$. The fixed point problem
is now associated with the map:
\begin{equation}
{\cal T}_{qT_{b}}\Phi_{n}(t)=\Phi_{n}(t) \label{osc_map}
\end{equation}
We cannot, however, develop the Newton iteration scheme in a
similar way as for mobile breathers. There is no longer any family
of oscillating breathers providing a good start point for the
continuation (they are intrinsic solutions of the nonintegrable
regime because they appear as the Peierls barrier emerges). The
way to obtain a good {\em ansantz} (as Cretegny and Aubry already
used to find mobile breathers in Klein-Gordon lattices 
\cite{CretegnyAubry}) is to perform a small perturbation of 
the static solution (pinned at a site $n$) with the depinning 
internal mode:
\begin{equation}
\Phi_{n}^{ansantz}=\Phi_{n}^{static}(\omega_{b})+\epsilon\delta\phi_{n}^{dep}
\end{equation}
The dynamics of the perturbed solution for small enough values of
$\epsilon$ shows the oscillating behavior expected and for large
enough values of $\epsilon$ the breather starts to move. Obviously
in both cases the motion finishes after a transient due to
radiation (they are not exact solutions). Tuning the parameter
$\epsilon$ we search for those oscillatory transients whose
$\omega_{osc}$ is resonant with the breather frequency $\omega_b$.
The transient is much more stable when the nonintegrable parameter
$\nu$ is very small, close to the A-L limit. We first search here
for a good initial guess for the method and then obtain the exact
solution of the map (\ref{osc_map}). Once the exact solution is
obtained for a small $\nu$, we can perform the continuation to
higher values in the same way as we did for mobile solutions. In
Fig. \ref{fig:Osc}.a we show the evolution of the amplitude of
oscillation as $\nu$ is increased from $0.05$ to $0.18$. The
amplitude of the oscillation is represented by the phase portrait
of the localization center of the breather defined as in
(\ref{loc_cent}). The continuation reflects that the amplitude of
the oscillation, for a fixed value of $\omega_{osc}$, grows with
$\nu$. In Fig. \ref{fig:Osc}.b the density plot of
$|\hat{\Phi}_{n}|^2$ is shown as a function of time, revealing the
oscillating pattern of the solution.

\begin{figure}[!tbh]
\begin{tabular}{cc}
\centerline{ 
{\bf (a)}
\resizebox{8cm}{!}{%
\includegraphics[angle=-0]{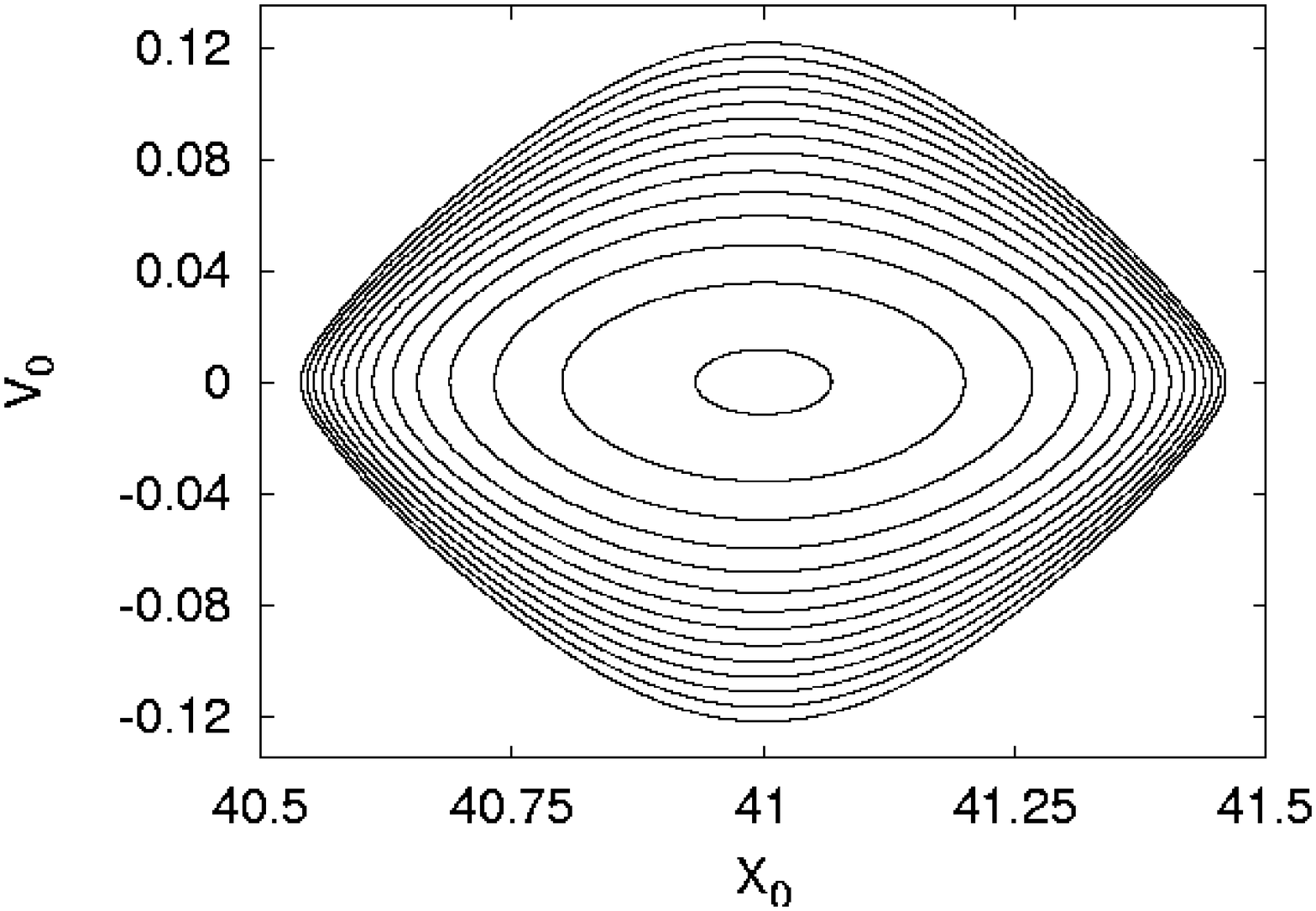}
}}
\\
\centerline{ 
{\bf (b)}
\resizebox{8cm}{!}{%
\includegraphics[angle=-0]{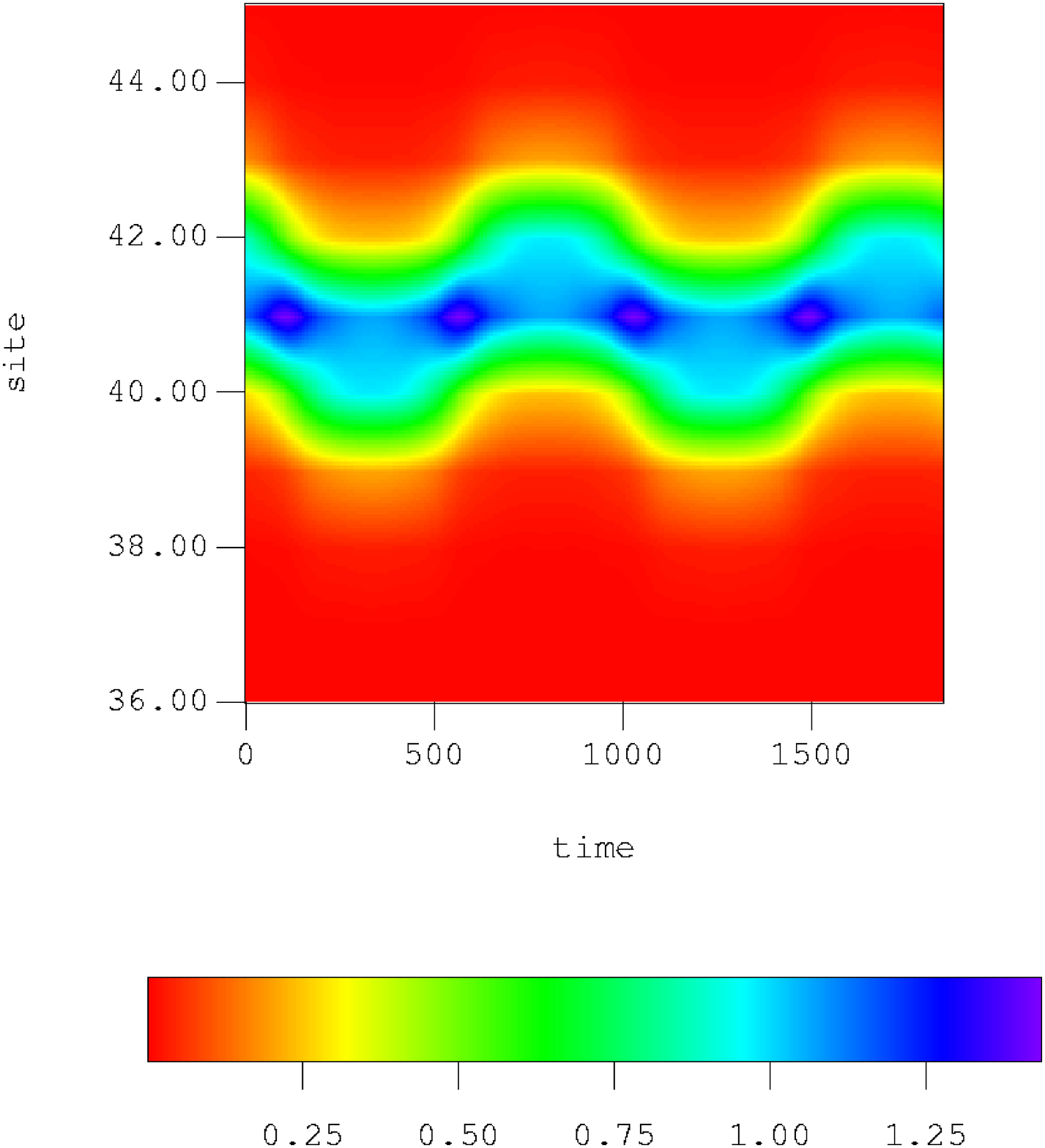}
}}
\end{tabular}
\caption{{\bf (a)} Evolution of the localization center $x_{0}$ of
an exact $(1/18)$-oscillating breather for different values of
$\nu$: $0.05$, $0.06$,.., $0.18$. The internal frequency is
$\omega_{b}=3.086$. The amplitude of the  oscillation of $x_{0}$ increases 
with $\nu$ revealing the nonlinear character of the motion for the highest
values of $\nu$. {\bf (b)} Density plot of the time evolution of 
$|\hat{\Phi}_{n}|^2$ for the above oscillating breather.} \label{fig:Osc}
\end{figure}

The existence of exact oscillating breathers is a consequence of
the existence of a Peierls barrier. The structure of these
solutions reveals the existence of a background (resonant with the
map) whose amplitude grows as $\nu$ (and consequently the
amplitude of oscillation) is increased. This is the picture we
expected from the role played by the interaction background-core
in the energy balance during motion. The monotonous growing
behavior of the background versus the oscillation amplitude,
strongly suggests that if the amplitude of the former is increased
the solution will be able to translate steadily. This has been
checked by direct numerical integration, because no exact
solutions connecting the oscillating with the mobile ones can be
obtained due to the different maps employed to obtain both types
of solutions. However, the existence of a background in the exact
oscillating breather solutions and its behavior with the amplitude
of the breather oscillations are fully consistent with the
interpretation of the results obtained for the mobile solutions.

\subsection{Validity and limitations of particle perspective}

The most basic result of the perturbative collective variable
theories away from the integrable regime is the existence of 
a Peierls-Nabarro potential function of the core (collective 
variable) center. It expresses (in particle-like terms) that 
the breather position is no longer indifferent because 
the continuous translational invariance has been broken. 
From this also naturally comes the existence of oscillating 
breathers. We have seen how our numerics fully confirm
the qualitative validity of these predictions.

A further prediction concerns the phase portrait's transition
studied in \cite{Claude}. Despite the fact that our end of
continuation is correlated with the equipotential lines profile of
the numerical PN barriers, and the phase portrait transition is
also related to their sudden growth, no clear connection
(between transition and end of continuation) can be
established. The end of continuation is itself sensibly
interpreted as a numerical consequence of the sudden increases of 
the amplitude background, and does not imply neccesarily the 
existence of any global phase portrait transition.

However, in some respects the perturbative collective variable
theory is clearly incomplete: For example, it is unable to predict
the observed localized (core) instability bifurcation and the observed 
symmetry breaking transitions for $\nu<0$. These bifurcations could 
easily appear in a theory with (at least) two variables (a dimer) 
experiencing the Peierls-Nabarro potential, which would demand an 
improved perturbative ansatz. This improved ansatz must coincide in the
integrable limit with the A-L solution. One can use the 
numerical results to guide the construction of such an improved 
ansatz. In this respect, the following observation may be relevant. 
The parameter $\beta$ of the A-L solution determines both the 
amplitude and the width of the localized pulse. However, our numerical 
estimates of these breather characteristics for immobile breathers 
show clearly that, for fixed value of $\omega_b$, the breather width is
independent of $\nu$, while the amplitude varies with it. In other
words, away from integrability, width and amplitude of the
(immobile) breather are no longer a single collective variable.

Beyond any other limitation of the perturbative collective variable
theory, the background (an indispensable part of the exact
solution) is absent in the perturbative ansatz, and it cannot
appear later in that context. A complete theory of
(nonlinear Schr\"odinger) breather motion should somehow incorporate the
background in the ansatz itself. If correct, it should then predict
that the background amplitude grows from zero with the
nonintegrability parameter $\nu$, and (ideally) so on with all
the numerically observed behaviors. One possible way to develop 
the analytical approach could be to use the method presented 
in \cite{Boesch}. In this scheme, eq. (\ref{Energy_balance}) may 
play an important role, for it provides the energy balance governing 
the translational motion of the breather core. In other words, our 
results show that the core energy is not an invariant of motion 
and this requires the existence of a finely tuned background whose 
nonlinear interaction with the core compensate the core energy 
variations. We hope that the numerical work presented here will 
provide motivation for further analytical research.

\section{Conclusions and prospective remarks}

We have studied numerically the problem of mobility of nonlinear
localized solutions of the one-dimensional Nonlinear Schr\"odinger
lattice (\ref{Salerno}) using a regularized Newton method. This 
method allows us to continue the family of mobile Ablowitz-Ladik 
discrete breathers into the nonintegrable domain of model
parameters. We find that these solutions decay asymptotically, 
in space, to an excited lattice extended state (the background), 
whose amplitude vanishes at the integrable Ablowitz-Ladik limit. 
This component of the solution is unambiguously found to be a 
linear combination of nonlinear resonant plane waves whose amplitudes 
decay typically, in $k$-space, exponentially. The exponentially 
localized oscillation (the core) of the amplitude probability 
rides over this extended radiation state.

\begin{equation}
\hat{\Phi} = \hat{\Phi}_{{\mbox{ core}}} + \hat{\Phi}_{{\mbox
{backg}}} 
\label{core-bckg2}
\end{equation}

Numerically exact moving discrete breathers with an infinitely
extended tail of small amplitude were already observed in some 
cases for Klein-Gordon lattices with Morse potential by
Cretegny and Aubry \cite{CretegnyAubry}, however no investigation
of the background of these exact solutions is reported, so they 
were able only to "..suggest that generally a
strictly localized breather cannot propagate without radiating
energy". Our systematic study of the NLS lattice allows us to go
further by showing that the extended background (here fully
characterized) plays an important and subtle role in the
translational motion of the localized core. Indeed, it is an
indispensable part of the exact solution in the nonintegrable
regime. Exact mobile localization only exists over finely tuned
extended states of the nonlinear lattice. Mobile "pure" ({\em
i.e.} rest state background) localization must be regarded as very
exceptional \cite{FlachZolotaryuk}.

The high numerical accuracy of the method allows a precise Floquet
stability analysis of the moving discrete breather solutions. The
Floquet analysis reveals the type of instabilities (both localized
and extended) eventually experienced by the two-parameter family
of moving breathers. Some generic bifurcations in the space of
model parameters are thus identified. For negative values of the
nonintegrability parameter $\nu$, we find narrow regions where
the immobile breathers experience mirror symmetry-breaking
bifurcations and, simultaneously, the amplitude of the background
component of the mobile breather solutions decreases down to
almost negligible values.

The most relevant predictions of perturvative collective variable
theory are confirmed by our numerical results, which show the
existence of Peierls-Nabarro barriers to breather translational
motion. Furthermore, the existence of exact oscillating breather
solutions is numerically confirmed. They are found to contain an
extended background whose amplitude is typically much smaller than
for mobile breathers.

The Peierls-Nabarro barrier $E_{PN}$, computed from immobile
breathers, and the amplitude background of moving breathers are
found to be strongly correlated, which correctly suggests that the
background has a role in the energy balance required to overcome
the barriers to translational motion. This is also fully
consistent with the observations on the background amplitude
behaviour of spatially oscillating anchored breathers. Indeed,
assuming the simplest case of a monochromatic background, the
variation of the core energy during its translational motion is
exactly compensated by the variation of the energy of interaction
between the localized core and the background. Currently used
effective particle (collective variable) theories are thus seen as
intrinsically incomplete, because core energy is not an invariant
of motion. Any sensible improved approach must adopt equation
(\ref{core-bckg2}) as starting point for improved perturbative
ansatze, and we hope that our work will estimulate further
studies along these lines.

There are, at very different levels, several open questions to
further research. From a technical point of view, it is important
to analyze carefully the irrational limit $p/q \rightarrow
\sigma$, of the solutions. In particular, in this limit the number
of resonant plane wave branches tends to a continuum and one could
(or not) expect that exponential localization in the reciprocal
lattice persists in that limit. This can be addressed numerically,
though systematic investigations may require some efforts in
optimizing the time efficiency of current numerical schemes.

An important issue regarding applications is the phenomenology of
multibreather states. In particular, studies on collisions of a
pair of breathers may find in this study of exact mobility a
useful reference in order to deal with the complexities that emerge
from the many time-length scales involved in these physically
relevant phenomena. Much simpler multibreather states, {\em e.g.}
train-like chains of (moderately) separated moving breathers could
also be investigated. Not least, the perspective and results
presented here may be of some interest to studies of the effects
of coupling to (nonthermal and/or thermal) radiation baths in the
breather and multibreather states of nonlinear lattices
\cite{Rasmussen} and the practical manipulation and patterning of 
localized "hot spots" by external fields \cite{CaiBisGronSal}.


\begin{acknowledgments}

We acknowledge interesting discussions with P. Kevrekidis, B. Malomed,
S. Flach, F. Falo, J.L. Garc\'{\i}a-Palacios, P.J. Mart\'{\i}nez,
J.-A. Sepulchre and G.P. Tsironis. We are grateful to the organisers
of the workshop ``{\em Intrinsic localized modes and discrete breathers
in nonlinear lattices}'' at Erice, where this collaboration was
initiated. Financial support came from LOCNET HPRN-CT-1999-00163,
MCyT (project BFM2002-00113), DGA and BIFI. 
J.G.-G. aknowledges financial support of the MECyD trough a FPU
grant. Work at Los Alamos performed under the auspices of the US DoE.

J.G.-G. would like to dedicate his contribution to this work to the
memory of his grandfather who recently disappeared.

\end{acknowledgments}

\bigskip\bigskip


\ifx\undefined\allcaps\def\allcaps#1{#1}\fi\ifx\undefined\allcaps\def\allcaps#%
1{#1}\fi



\begin{thebibliography}{10}
\bibitem{note}
However, in reference \cite{FlachZolotaryuk} the authors find
particular on-site potentials for Klein-Gordon lattices for which
moving kinks and pulses do not posses an infinite tail, though
immobile breathers do experience a Peierls-Nabarro barrier.
\end{thebibliography}
\end{document}